\documentclass[english,aps,prd,superscriptaddress,preprintnumbers,nofootinbib,preprint]{revtex4}
\usepackage[T1]{fontenc}
\usepackage[latin9]{inputenc}
\usepackage{verbatim}
\usepackage{hyperref}

\usepackage{empheq}
\usepackage{amssymb}
\usepackage{amsmath}
\usepackage{amsfonts}
\usepackage{multirow}

\usepackage{dcolumn}
\usepackage{graphicx}
\newcommand{\be}{\begin{equation}}
\newcommand{\ee}{\end{equation}}
\newcommand{\bea}{\begin{eqnarray}}
\newcommand{\eea}{\end{eqnarray}}

\newcommand{\lp}{\left(}
\newcommand{\rp}{\right)}


\usepackage{babel}
\usepackage{color}
\begin{document}
\global\long\def\order#1{\mathcal{O}\left(#1\right)}
\global\long\def\d{\mathrm{d}}
\global\long\def\P{P}
\global\long\def\amp{{\mathcal M}}
\preprint{IPPP/17/107, TTP17-056, MPP-2017-257}

\def \SS {S{\hspace{-5pt}}S}
\def\KIT{Institute for Theoretical Particle Physics, KIT, Karlsruhe, Germany}
\def\DUR{Institute for Particle Physics Phenomenology, 
Durham University, Durham, UK }
\def\MPI{Max Planck Institute for Physics, F\"ohringer Ring 6, 80805 M\"unchen, Germany}

\title{ 
NNLO QCD corrections to associated $WH$ production \\ and $H \to b \bar b$
decay 
}

\author{Fabrizio Caola}            
\email[Electronic address: ]{fabrizio.caola@durham.ac.uk}
\affiliation{\DUR}

\author{Gionata Luisoni}
\email[Electronic address: ]{luisonig@mpp.mpg.de}
\affiliation{\MPI}

\author{Kirill Melnikov}            
\email[Electronic address: ]{kirill.melnikov@kit.edu}
\affiliation{\KIT}

\author{Raoul R\"ontsch }            
\email[Electronic address: ]{raoul.roentsch@kit.edu}
\affiliation{\KIT}

\begin{abstract}
We present a computation of the next-to-next-to-leading order (NNLO) 
QCD corrections to the production of 
a Higgs boson in association with a $W$ boson at the LHC and 
the subsequent decay of the Higgs boson into a $b \bar b$ pair, treating 
the $b$-quarks as massless.  
We consider various kinematic distributions and find significant 
corrections to observables that resolve the Higgs decay 
products. We also find that a cut on the transverse momentum of the  
$W$ boson, important for
experimental analyses, may have a significant impact 
on kinematic distributions and radiative corrections. 
We show that some of these effects can be adequately described by simulating 
QCD radiation in Higgs boson decays to 
$b$-quarks using parton showers. 
We also describe contributions to Higgs decay to a $b \bar b$ pair that 
first appear at NNLO  and that were not considered in 
previous fully-differential computations. 
The calculation  of NNLO QCD corrections to production and decay sub-processes
is carried out within the nested soft-collinear subtraction scheme presented 
by some of us earlier this year. We demonstrate that this subtraction scheme 
performs very well, allowing a computation of the coefficient of the second order
QCD corrections at the level of a few per mill.

\end{abstract}

\maketitle

\section{Introduction} 

Production of the Higgs boson in association with the $W$ 
boson $pp \to WH$ plays an 
important role in  Higgs physics explorations 
at the LHC \cite{Aad:2014xzb,ATLAS:2016pkl,Chatrchyan:2013zna,Aaboud:2017xsd}.
For example, it provides 
direct access to the $HWW$ coupling, which is completely fixed  by 
the gauge symmetry of the Standard Model (SM) but may 
receive new contributions in its extensions.
 The 
$WH$ associated production is known to provide important constraints on such
anomalous couplings,
see e.g. Ref.~\cite{Ellis:2013ywa}. 
Furthermore, as was pointed out in  Ref.~\cite{Butterworth:2008iy}, 
by selecting  Higgs bosons with relatively 
high transverse momenta, it is possible  
to identify and study the decay of a Higgs boson into a $b \bar b$ pair 
with high efficiency. The associated $WH$ production then becomes sensitive to the value 
of the bottom quark Yukawa coupling which currently is only constrained
to within a factor of two relative to its SM value
\cite{Aaboud:2017xsd,couplATLCMS}.

The importance of associated $WH$  production   inspired a  large number of computations
of higher-order QCD and electroweak (EW) corrections
   to this process. 
The next-to-leading order (NLO) QCD and EW corrections to $pp \to WH$ were computed in 
Refs.~\cite{than} and \cite{Ciccolini:2003jy,Denner:2011id}, respectively. 
NLO QCD and EW fixed order computations were subsequently matched to parton
showers in Refs.~\cite{NLOPS,NLOJPS}.
The inclusive next-to-next-to-leading order (NNLO) QCD corrections 
to $pp \to WH$  can be deduced~\cite{Brein:2003wg} from the  NNLO QCD corrections 
to the Drell-Yan process 
$pp \to W^*$ computed in Refs.~\cite{Hamberg:1990np,Harlander:2002wh}.
Additional  NNLO QCD effects that distinguish associated production from  the 
Drell-Yan  process  originate from diagrams where the Higgs boson is emitted 
by loops of  top quarks; these effects  were computed in Ref.~\cite{Brein:2011vx}
in the large top mass approximation. 
The numerical program {\tt VH@NNLO}, which allows high-precision computations of the 
inclusive cross section of 
associated Higgs boson production, was developed in Ref.~\cite{Brein:2012ne}.

Fully-differential NNLO QCD results for associated $WH$ production were obtained 
in  Refs.~\cite{Ferrera:2011bk,Campbell:2016jau} using slicing techniques. The 
NNLO calculation of Ref.~\cite{Ferrera:2011bk} was matched to a parton shower 
in  Ref.~\cite{Astill:2016hpa}. 
NLO QCD corrections to $H \to b\bar b$ decay were combined
with NNLO QCD corrections to the $pp\to WH$ production process in Ref.~\cite{Ferrera:2013yga},
in the limit of a vanishing $b$-quark mass, and in Ref.~\cite{Campbell:2016jau},
retaining the full $m_b$ dependence. Recently, the computation of Ref.~\cite{Ferrera:2013yga} was extended 
\cite{Ferrera:2017zex} to include the NNLO QCD 
corrections to $H \to b \bar b$ decay computed earlier 
in Ref.~\cite{DelDuca:2015zqa} (see also Ref.~\cite{Anastasiou:2011qx}), in the limit 
$m_b\to 0$. Very large  effects, apparently  caused by an improved treatment of 
radiative corrections in the  decay $H \to b \bar b$, 
were  found for some  kinematic distributions. 

The purpose of this  paper is to repeat the computation 
of Ref.~\cite{Ferrera:2017zex}.  There 
are several reasons for doing so. First, it is important to check the 
appearance of  large effects when QCD corrections to decays are included. 
Also, we note that some peculiar contributions to Higgs decay to a 
$b \bar b$ pair  
that appear at NNLO QCD for the first time  were not considered in the computations of 
Refs.~\cite{Anastasiou:2011qx,DelDuca:2015zqa} and we discuss them here.  

Second, the type of distributions for which large QCD corrections 
were found in Ref.~\cite{Ferrera:2017zex}   
are typically pathological at leading 
order. For example,  kinematic requirements can result in certain 
regions of phase space only being populated at NLO. 
In these kinematic regions, the NNLO computations provide next-to-leading 
order corrections so that moderately large effects are not too surprising. 
In addition, severe cuts on the final state particles imply the appearance of 
kinematic boundaries that may cause genuine large effects that signal poor 
convergence of perturbation theory. In general, 
many of these effects are driven by parton emissions and may be 
properly described by parton showers. 
It is then interesting to check to what extent the large radiative 
corrections found in Ref.~\cite{Ferrera:2017zex} can be described by a 
parton shower applied to $H \to b \bar b$ decay.

Finally, we perform the computation using the local subtraction scheme  
described recently in Ref.~\cite{Caola:2017dug}.
This scheme is  an extension of the 
original sector-improved residue subtraction scheme developed 
in Refs.~\cite{Czakon:2010td,Czakon:2011ve}.
As we already mentioned, all  previous computations of $WH$ production 
at NNLO QCD  were  performed using variants of 
the slicing method and it is interesting to 
perform the computation using fully local subtractions. 

The remainder of the paper is organized as follows. 
In Section~\ref{sect:basics} 
we briefly review the computational scheme of  Ref.~\cite{Caola:2017dug} 
with an eye on its application to the production process $pp \to WH$. 
In Section~\ref{sect:WHprod} we illustrate the performance of the subtraction
scheme by showing numerical results for 
NNLO QCD corrections to the $pp \to WH$ process, 
treating the $H\to b \bar b$ decay at leading order. 
In Section~\ref{sect:Hbbdec}
we  discuss the generalization   
of the scheme of  Ref.~\cite{Caola:2017dug}  to 
the  decay process $H \to b \bar b$, and
point out  differences between the production and decay cases.
We also present numerical results for the 
NNLO QCD corrections to the $H\to b \bar b$ decay process, to illustrate
the performance of the subtraction scheme in this case as well. 
Finally,  we discuss the phenomenology of the process  $pp \to W(l\nu) H (b \bar b)$,
consistently including NNLO QCD corrections to both production and decay. We present 
numerical results for cross sections and selected distributions 
in Section~\ref{sect:fullpheno} 
and compare them with the approximate treatment of QCD  corrections 
to $H \to b \bar b$ decay using a parton shower in Section~\ref{sec:partsh}.
We conclude  in Section~\ref{sect:concl}.

\section{Basics of NNLO QCD computations within the nested soft-collinear
subtraction framework}
\label{sect:basics}

The goal of this  Section is to review the subtraction scheme for NNLO QCD
calculations~\cite{Caola:2017dug}.
We consider the collision of two partons and ask for the fiducial volume 
cross section defined by an infra-red and collinear-safe 
observable ${\cal O}$. The fiducial 
cross section is schematically written as 
\be
\sigma_f({\cal O}) = \sum \limits_{X}^{}\int {\rm dLips}(\{p_X \}) 
 |{\cal M}|^2(\{p_X\}) {\cal O}(\{p_X\}),
\label{eq:sigma}
\ee
where ${\rm dLips}$ is the Lorentz-invariant phase space and
$\cal M$ is the amplitude for the process $X$. In Eq.~(\ref{eq:sigma}),
final states $X$ of  increasing multiplicity have to be 
included to arrive at a high-order result for $\sigma_f$. 
In our case, the leading-order computation  includes partonic 
processes of the type $q \bar q' \to WH$ followed by the decays 
$H \to b \bar b$ and $W \to l \bar \nu$. 
Both the production and the $H \to b \bar b$ decay processes 
are affected by QCD corrections.  In 
this Section we focus on the QCD corrections to $WH$ associated 
production and consider Higgs decay in the leading-order approximation.

We note that the NNLO QCD corrections to inclusive $WH$ production are known 
since long ago \cite{Brein:2003wg,Brein:2011vx,Brein:2012ne}. 
The challenge for an exclusive computation 
is to extract the soft and collinear divergences from, say, a matrix 
element squared with two additional final state partons 
relative to the leading-order matrix element, while avoiding integration over  
momenta   of   partons that can get resolved. 

 At next-to-leading  order, an understanding 
of how to do this in full generality using  both slicing and subtraction 
methods  was achieved  more than twenty years ago  
\cite{nloslice,Catani:1996vz,Frixione:1995ms,Frixione:1997np}.
Unfortunately, the generalization of these methods to NNLO proved to be difficult 
and required significant effort. This 
effort started to pay off in the past two to three  years,  and 
a large number of fully-differential NNLO QCD results for 
important LHC processes has been 
obtained using different computational 
methods \cite{ant,Czakon:2010td,Czakon:2011ve,Czakon:2014oma,njet1,njet2,grazzi,P2B,colorful}.

One of these methods, the so-called  sector improved residue-subtraction 
scheme, was developed in Refs.~\cite{Czakon:2010td,Czakon:2011ve} (see 
also \cite{Boughezal:2011jf,Czakon:2014oma} for related work). 
Recently, it was shown~\cite{Caola:2017dug}
how to modify the original formulation of the 
method by exploiting the fact that in QCD soft and collinear singularities 
are not entangled.  This allows one to closely follow the 
so-called FKS subtraction scheme \cite{Frixione:1995ms,Frixione:1997np},
developed for NLO QCD computations, and perform the required 
soft and collinear subtractions  in a nested way~\cite{Caola:2017dug}.  
As a consequence, the computational framework  
becomes very transparent and, as we show below, numerically efficient. 

We will illustrate the main idea of Ref.~\cite{Caola:2017dug} 
by considering 
the double-real emission contribution, taking the 
process $q(p_1)  \bar q'(p_2) \to WH +g(p_4) g(p_5)$ as an example. Final states with lower 
multiplicity can be treated along the same lines although the details 
can be slightly different.  Schematically, we write the corresponding  cross section as 
\be
\sigma_f^{gg}({\cal O}) = \int [{\rm d} g_4] [{\rm d}g_5] \theta(E_4-E_5)
F_{LM}(1,2,4,5)
= \langle F_{LM}(1,2,4,5) \rangle, 
\label{eq2}
\ee
where 
\be
F_{LM}(1,2,4,5) = 
\int {\rm dLips}(p_1+p_2-p_4-p_5 \to W+H)) 
|{\cal M}|^2(\{p\}) {\cal O}(\{p\}),
\ee
and
\be
[{\rm d} g_i ] = \frac{{\rm d}^{d-1} p_{g_i}}{(2\pi)^{d-1} 2E_{g_i}}
\theta(E_{\rm max} - E_{g_i}) 
\ee
is the phase-space element for a gluon,  supplemented with 
a $\theta$-function that ensures that the gluon energy is bounded 
from above.  Note that 
we introduced the energy ordering of gluons in Eq.~(\ref{eq2}) 
to remove the $1/2!$ identical particles factor. 

Our goal is to extract  singularities from Eq.~(\ref{eq2}). 
These singularities can occur in several ways. 
For example, the so-called double-soft singularity arises 
if the energies of the two gluons vanish simultaneously. 
A single-soft singularity appears if $E_5$ vanishes  at fixed 
$E_4$. Note that due to the  energy ordering in Eq.~(\ref{eq2})  the opposite limit 
 ($E_4 \to  0$ at fixed $E_5$) cannot occur. In addition to these soft 
singularities, there are also collinear singularities that occur when 
the gluons 
are  emitted along the direction of the incoming quark, incoming 
anti-quark or if they are emitted collinear to each other.  

We need to extract all these singularities in an unambiguous way.  We begin with soft singularities. 
We write 
\be
\sigma_f^{gg}({\cal O})  = \langle F_{LM}(1,2,4,5) \rangle 
= 
\langle \SS F_{LM}(1,2,4,5) \rangle 
+
\langle ( I - \SS) F_{LM}(1,2,4,5) \rangle,
\label{eq5}
\ee
where $\SS$ is an operator that extracts the double-soft\footnote{Here we define
the double-soft limit as $E_4 \to 0, E_5 \to 0$ at fixed
$E_5/E_4$.} singularity from 
$F_{LM}$. When the operator $\SS$ acts on $F_{LM}$, 
it  removes the four-momenta of the gluons from  both 
the energy-momentum conserving $\delta$-function inside ${\rm dLips}$ and the 
observable $\mathcal O$,  and extracts the leading singular behavior from the matrix element
squared. The result is well known
\be
\SS F_{LM}(1,2,4,5) = g_s^2\;  {\rm Eik}(1,2,4,5) \; F_{LM}(1,2),
\label{eq6}
\ee
where ${\rm Eik}(1,2,4,5)$ is the square of the eikonal factor derived in 
Ref.~\cite{grazzi}. It is also given in   Ref.~\cite{Caola:2017dug} 
using the same notation as we use in this paper. 

We deal with the  two terms on the right-hand side of Eq.~(\ref{eq5})  
in different ways. In the first term  the hard matrix element decouples 
thanks to Eq.~(\ref{eq6})  and only the eikonal factor needs to be integrated 
over the two-gluon phase-space.  This integral was performed numerically 
in Ref.~\cite{Caola:2017dug}.  The second term in Eq.~(\ref{eq5}) has its 
double-soft divergences regularized. However, both the $E_5 \to 0$ 
divergence at fixed $E_4$ as well as the collinear divergences 
are still present there.  To take care of them, we repeat the 
procedure and subtract the  $E_5 \to 0$ singularities 
at fixed  $E_4$.  We call the corresponding 
operator $S_5$ and write 
\be
\langle ( I - \SS) F_{LM}(1,2,4,5) \rangle  = 
 \langle (I - \SS) (I - S_5) F_{LM}(1,2,4,5) \rangle
+\langle S_5 (I - \SS) F_{LM}(1,2,4,5) \rangle .
\label{eq7}
\ee
The operator $S_5$ acting on $F_{\rm LM}(1,2,4,5)$ removes the gluon 
$g_5$ from the phase space and the observable and extracts the leading 
singularity 
\be
S_5 F_{LM}(1,2,4,5)  = 
\frac{g_s^2}{E_5^2}
\left [ (2 C_F - C_A) \frac{\rho_{12}}{\rho_{15} \rho_{25}}
+ C_A \left ( \frac{\rho_{14}}{\rho_{15} \rho_{45}} + \frac{\rho_{24}}{
\rho_{25} \rho_{45} } \right ) \right ] F_{LM}(1,2,4).
\label{eq8}
\ee
We use the notation  $\rho_{ij} = 1 - \cos \theta_{ij}$ 
in Eq.~(\ref{eq8}), where $\theta_{ij}$ is the relative 
angle between partons $i$ and $j$. Among the two terms on the 
right hand side in Eq.~(\ref{eq7}), 
the first has only collinear divergences and the 
second has a simplified (i.e. independent of $g_5$) 
matrix element. Therefore, the integration over the energy and 
emission angles of the gluon $g_5$ can be performed in this term. 
The remaining matrix element for the 
process $q \bar q' \to WH + g_4$ can then be treated similarly
to a normal NLO computation.

The procedure  continues with collinear subtractions  that are again applied 
to the terms on the right hand side in Eq.~(\ref{eq7}) on top  of the soft subtractions 
shown there. However, an additional step,  similar to the 
energy ordering in Eq.~(\ref{eq2}), is required. Indeed, we need to
further split the phase space 
into sectors such that in each of them only a particular type
of collinear singularity can occur.  

There are two major ingredients to this phase space splitting. First, we partition 
the phase space into two double-collinear partitions and two triple-collinear 
partitions. In the two double-collinear partitions, the gluons can only 
have singularities if $\vec p_4 || \vec p_1, \vec p_5 || \vec p_2$,  
or if $\vec p_4 || \vec p_2, \vec p_5 || \vec p_1$, respectively.  In the two 
triple-collinear partitions, singularities appear if $\vec p_1 || \vec p_4 || \vec p_5$ 
or  if $\vec p_2 || \vec p_4 || \vec p_5$, respectively. We note that in 
the two latter cases the singularities  can also appear if $\vec p_4 || \vec p_5$.

The contributions of the double-collinear partitions can be computed  right away 
since all singular limits are uniquely established. The situation is more 
complex for the triple-collinear partitioning where this is not the case. 
Indeed, in triple-collinear configurations we need to consider the two cases
of the gluons being either close or well-separated in rapidity. 
To this end, we further partition the phase space into four  sectors.
Taking as an example the $\vec p_1 || \vec p_4 || \vec p_5$  partitioning, 
we introduce four sectors according to the following formula 
\be
\begin{split}
1 = \theta \left (\rho_{51} < \frac{\rho_{41}}{2} \right ) 
+ \theta \left (\frac{\rho_{41}}{2} < \rho_{51} < \rho_{41}  \right ) 
+ \theta \left (\rho_{41} <  \frac{\rho_{51}}{2}  \right ) 
+ \theta \left ( \frac{\rho_{51}}{2} <  \rho_{41} < \rho_{51} \right ). 
\end{split}
\ee
Note that this splitting is largely arbitrary. 
The important point is that in each of the four sectors only a well-defined type
of singular collinear limit can occur; by choosing an appropriate parametrization, 
 these singularities 
can be  resolved and isolated. The nested subtraction of these collinear 
limits can then be performed, similar to what we discussed in connection 
with the soft limits. A convenient phase-space parametrization for each of the 
four sectors  can be found in  Ref.~\cite{Czakon:2010td}. 

A detailed discussion of  this approach  
can be  found in Ref.~\cite{Caola:2017dug} which an interested reader should consult. 
Below, we list a few aspects of the current computation that go beyond that reference. 
\begin{itemize}

\item We extend the computation  of Ref.~\cite{Caola:2017dug} 
by  including  the 
$qg \to WH + q'g$ partonic channel. The difference with the quark-anti-quark
annihilation channel is that the quark-gluon channel appears for the first time at NLO 
and, therefore, to obtain the result relevant for the  NNLO computation, 
we only need to include 
one-loop corrections to this  channel and consider one additional 
gluon in the final state.  There are no 
conceptual differences with the computations 
described in Ref.~\cite{Caola:2017dug} and, similarly to that  
reference,  compact formulas 
are  obtained  for the NNLO contribution of the quark-gluon channel. In our 
implementation, we used a slightly different parametrization of the phase space
compared to Ref.~\cite{Caola:2017dug}
making use of the fact that there are no single-soft singularities related
to quark emission. 

\item We compute all the channels with an additional quark-anti-quark 
 pair in the final  state $ q \bar q' \to WH + q_1  \bar q_2$. 
If the quark-anti-quark 
pair comes from gluon splitting, the corresponding process 
has a double-soft singularity that is different from the  
one described above; the integral of the respective eikonal 
factor has to be computed anew. 

\item We compute the contribution of the 
$gg \to WH + q \bar q'$ channel. This channel has  simple 
collinear divergences and  their extraction is straightforward.

\item We include the NNLO contributions to the associated production 
$pp \to WH$  where the Higgs boson is emitted from a 
loop of virtual top quarks. These include two-loop corrections 
to the $q\bar{q} \to WH$ partonic process, as well as one-loop corrections
to the $q\bar{q} \to WH+g$ process. These finite contributions
were first computed in Ref.~\cite{Brein:2011vx}, where they were referred to as 
$V_I$ and $R_I$, respectively. We take the amplitude for $V_I$ in an expansion 
in  $1/m_{\rm top}$ and the amplitude for $R_I$ with the exact mass
dependence from Refs.~\cite{Brein:2011vx,Campbell:2016jau}.

\end{itemize}

\begin{figure}[t]
\centering
\includegraphics[width=0.485\textwidth]{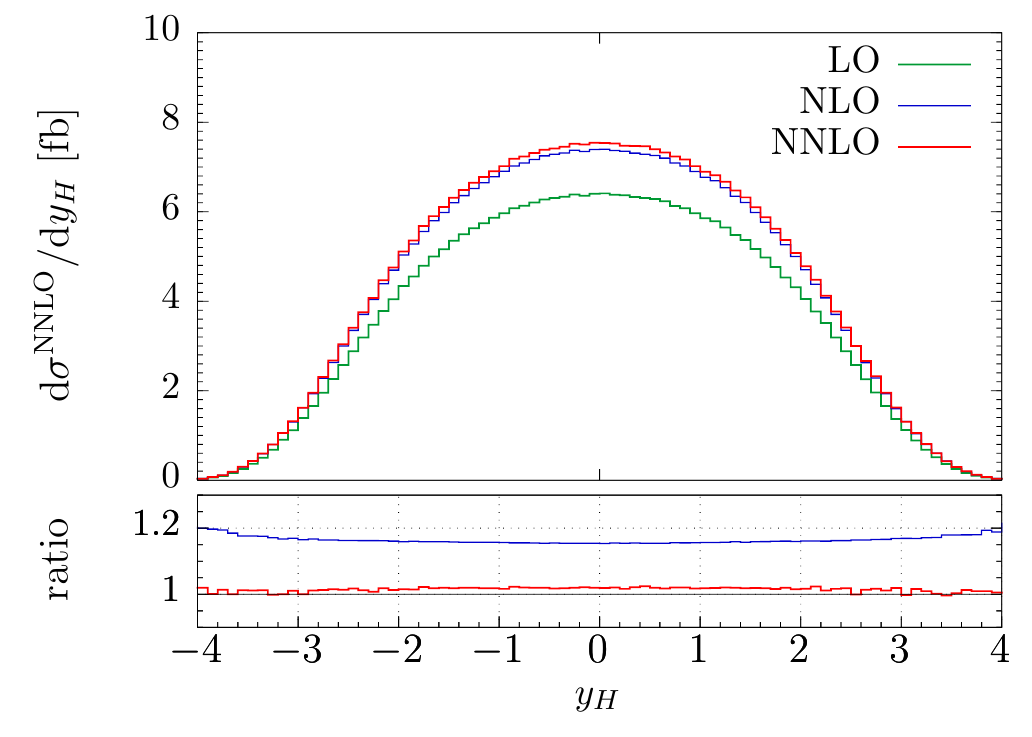}
~~
\includegraphics[width=0.485\textwidth]{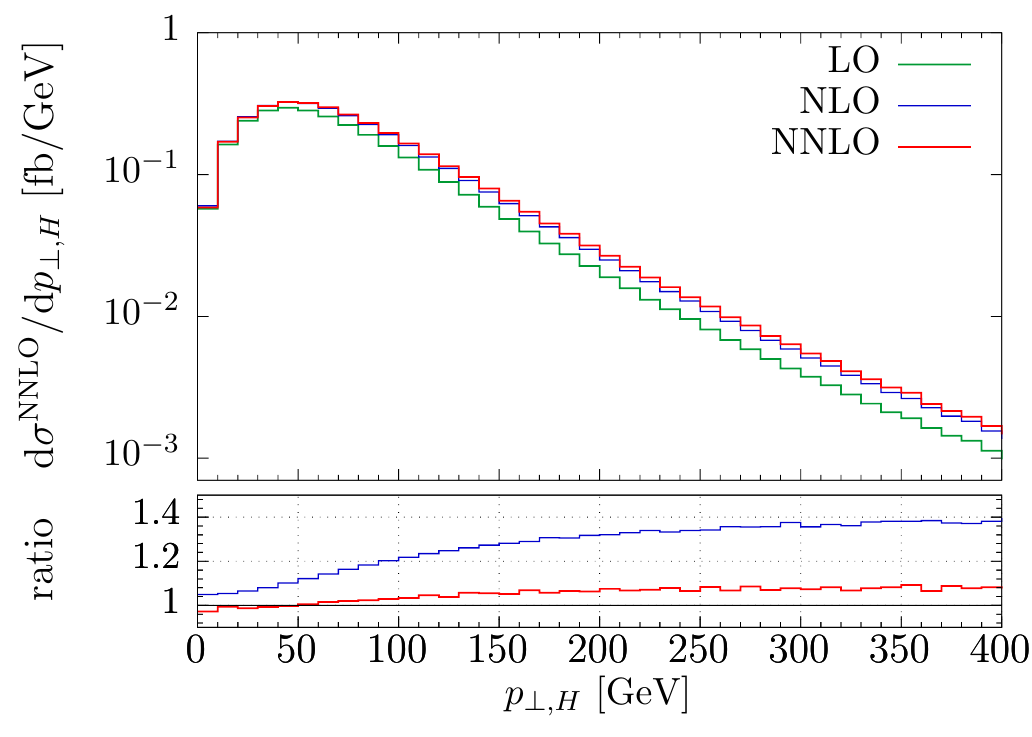}
\caption{Results for the rapidity and the transverse momentum 
distributions of the Higgs boson. 
Upper panes -- 
results in consecutive  orders of perturbation theory. 
Lower panes -- ratios of NLO to LO and NNLO to NLO.
The renormalization and factorization scales are set   
to  $\mu = M_{WH}$.
 In this plot, LO,  NLO and NNLO results
are all computed with NNLO PDFs,  see text for detail.}\label{WHplots}
\end{figure}

\section{ Associated $WH$ production} 
\label{sect:WHprod}

In this Section we present results of the computation 
of the NNLO QCD corrections to the process $pp \to W(l\nu) H(b\bar b)$
at the center-of-mass energy  $\sqrt{s} = 13~{\rm TeV}$. 
We compute  LO, NLO and NNLO 
cross sections and distributions {\it always} 
 using NNPDF3.0 NNLO parton distribution 
functions~\cite{Ball:2014uwa}. 
We use the numerical value of the strong coupling constant
provided by the PDF set, with $\alpha_s(m_Z)=0.118$.

The Higgs and the $W$ boson masses 
are taken to be $125~{\rm GeV}$ and $80.398~{\rm GeV}$,  respectively. 
The mass of the top quark is set to $173.2~{\rm GeV}$. The decays of the 
Higgs boson and of the $W$ boson are
 included in the narrow width approximation. In this Section, we consider
the $H\to b\bar b$ decay at LO only. 
The width of the $W$ boson is $\Gamma_W = 2.1054~{\rm GeV}$.
The Fermi constant is $G_F=1.16639 \times 10^{-5}~{\rm  GeV}^{-2}$, and  we take   
$\sin^2 \theta _W =0.2226459$  as the sine squared of  the  weak mixing angle.
We also approximate  the CKM matrix by an identity matrix.
For the decay of the Higgs boson, we take the $b$-quark Yukawa coupling to be
$y_b = \sqrt{2} \; m_b(m_H)/v = 0.0176$, which corresponds to $m_b(m_H) = 3.07~{\rm GeV}$. 
We consider only the leading term in $m_b$, which at this order corresponds to 
treating $b$-quarks as massless particles  but with a non-vanishing Yukawa coupling. 
Finally, the Higgs boson width is taken to be $\Gamma_H = 4.165~{\rm MeV}$.
 
We employ dynamic renormalization and factorization scales 
that we take to be proportional to the invariant mass of the 
$WH$ system $M_{WH}$. We compute the NNLO QCD corrections 
for  three values of the scales  
$\mu = M_{WH}/2$, $\mu =  M_{WH}$ and $\mu = 2 M_{WH}$, while keeping
the scale of the $b$-quark Yukawa coupling fixed to $m_H$.

\begin{table}[t]
\begin{center}
\scalebox{0.85}{
\begin{tabular}{|c|c|c|c||c|c|c|}
\hline
& \multicolumn{3}{c||}{$p p \to W^+ H \to l^+ \nu b \bar b$} &
  \multicolumn{3}{c|}{$p p \to W^- H \to l^- \bar \nu b \bar b $}\\
\cline{2-7}
& $\mu = M_{WH}/2$ & $\mu = M_{WH}$ & $\mu = 2 M_{WH}$ 
& $\mu = M_{WH}/2$ & $\mu = M_{WH}$ & $\mu = 2 M_{WH}$\\
\hline
$\sigma_{\rm LO}~({\rm fb})$ & 44.340(5) & 45.748(6) & 46.834(6) &
28.207(5) & 29.158(5) &  29.890(5) 
\\
\hline
$\sigma_{\rm NLO}~({\rm fb})$ & 53.475(6) & 53.286(6) & 53.284(7) &
33.867(5) & 33.773(5) & 33.805(5)
\\
\hline
$\sigma_{\rm NNLO}~({\rm fb})$ & 54.498(13) & 54.401(18) & 54.378(10)
& 34.452(5) & 34.402(5) & 34.397(5) 
\\
\hline
\end{tabular}}
\caption{Results for $pp \to W^+H \to l^+ \nu b\bar{b} $ (left) and 
$p p \to W^- H \to l^- \bar \nu b \bar b$ (right).
Higgs boson emissions off the top quark loops
are included.  Higgs decays are accounted for at the
LO approximation.
See text for details.
\label{tab:inclusive}
}
\end{center}
\end{table}
   
We report our results for $W^+$ and $W^-$ production in Table~\ref{tab:inclusive}.
NLO QCD corrections increase the leading 
order cross section by about 15\%; the 
NNLO  QCD corrections increases the NLO cross sections 
by an additional 2\%.  We note that  the 
scale dependence of the NNLO  cross sections is below a percent.
Therefore, it is both completely negligible and unlikely to be a 
reliable estimate of the actual theory uncertainty. This issue
has been discussed at length in Ref.~\cite{deFlorian:2016spz}, and we do not comment on it here. 
Ratios of $W^+$ and $W^-$ cross sections stay close to $1.57-1.58$, 
independent of both the order of perturbation theory and the choice of the factorization 
and the renormalization scales. 
We have cross-checked all these numbers against {\tt VH@NNLO}~\cite{Brein:2012ne} and 
found perfect agreement.

As the 
next step, we study the NNLO corrections in more detail, focusing on the case of $W^-H$ 
production.\footnote{Results for $W^+H$ production show a similar qualitative behavior.}
The  NNLO QCD {\it contributions} to the $W^-H$ production cross section without the finite 
contributions $R_I$ and $V_I$ describing Higgs boson emission off a 
 top quark loop read 
\be
\delta \sigma_{\rm NNLO}^{\rm no~top~loops}  = 
\{0.0937(7), 0.2193(7), 0.2464(7)\}~{\rm fb},
\ee
for the renormalization and factorizations scales  $\mu = \{M_{WH}/2,M_{WH},2 M_{WH}\}$  respectively.
Note that the numerical integration error on the NNLO \emph{coefficients}  is just a few  per mill.
Also in this case, full agreement with Ref.~\cite{Brein:2012ne} was found. 
The fact that our computational method is capable of 
delivering results at this level of numerical precision for the NNLO QCD coefficients 
has already been noticed in the calculation reported in 
Ref.~\cite{Caola:2017dug}.  However, since the calculation 
of Ref.~\cite{Caola:2017dug} was performed for a simplified case, 
it is gratifying to see 
that this feature  persists in a more complex  situation  where  all the 
different partonic channels are included in the calculation,  and  
significant  numerical 
cancellations between their contributions occur.

\begin{figure}[t]
\centering
\includegraphics[width=0.485\textwidth]{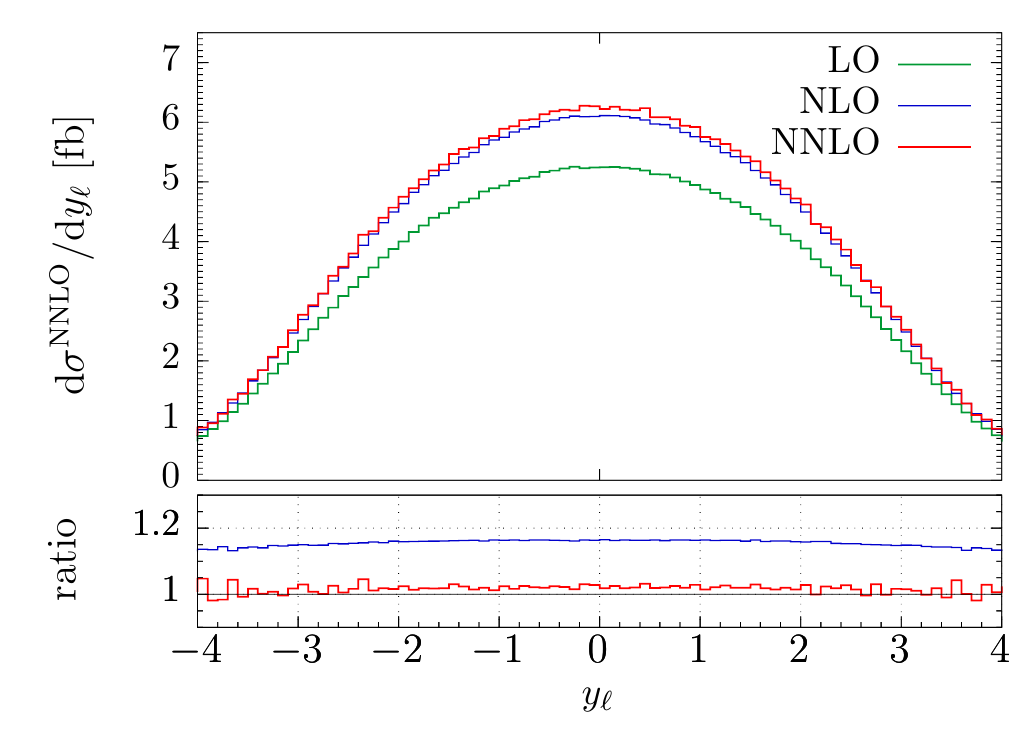}
~~
\includegraphics[width=0.485\textwidth]{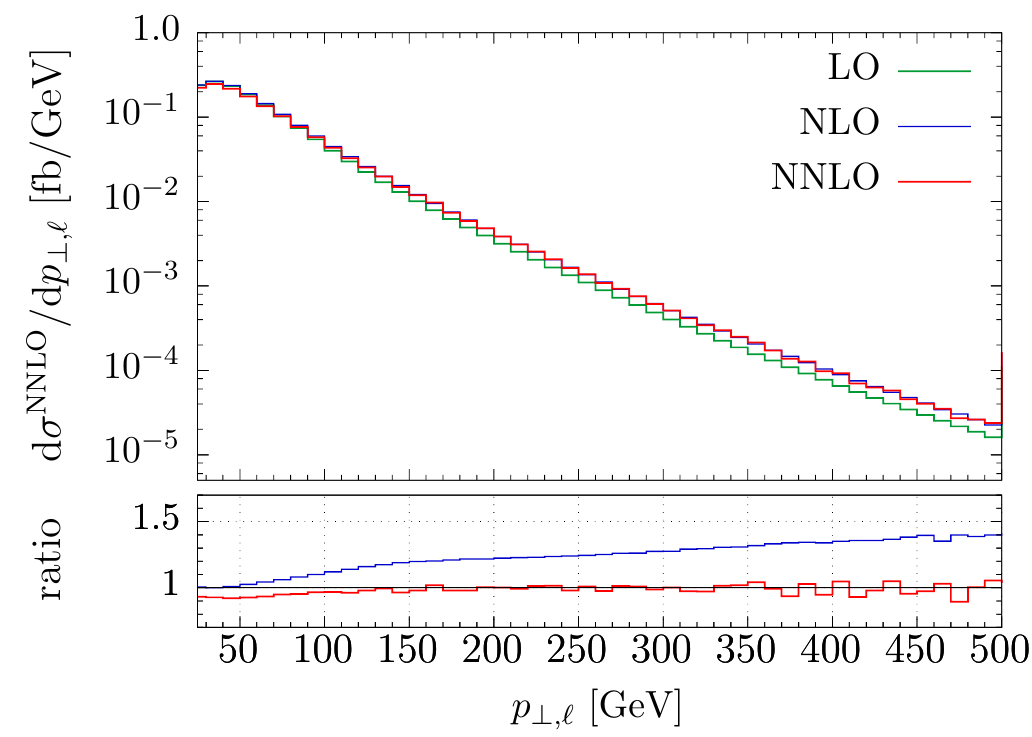}
\caption{Results for rapidity and transverse momentum 
distributions of the charged lepton from the decay of a $W^-$ boson.  
  Upper panes --
results in consecutive  orders of perturbation theory. 
Lower panes --  ratios of NLO to LO and NNLO to NLO. 
The renormalization and factorization scales are set   
to  $\mu = M_{WH}$. In this plot, LO, NLO and NNLO
results are all computed with NNLO PDFs, see text for detail.}\label{Wdecplots}
\end{figure}

NNLO QCD corrections to  kinematic distributions 
can also be computed with  a high degree of numerical stability. 
In Fig.~\ref{WHplots} we display the Higgs boson rapidity and 
transverse momentum   distributions in consecutive orders of QCD perturbation 
theory, for the case of $W^-H$ production. 
In the lower panels  of Fig.~\ref{WHplots} we also display 
ratios of NLO to LO and NNLO to NLO distributions. 
In Fig.~\ref{Wdecplots} we show results for the rapidity and 
the transverse momentum distributions of the charged lepton from 
the decay of the $W$ boson.  The numerical stability of these computations is clearly 
very good.

 \section{Higgs decay to a pair of bottom quarks} 
\label{sect:Hbbdec}

In this Section, we discuss a fully exclusive computation of NNLO QCD 
corrections to Higgs boson decay to a $b \bar b$ pair. 
Such computations were performed in 
Refs.~\cite{Anastasiou:2011qx,DelDuca:2015zqa}. 
Unfortunately, both of these  references 
did not  consider an interesting subtlety related 
to this decay that we will explain first. 

We consider the Standard Model Lagrangian, integrate out the top quark and 
neglect the  interaction of the Higgs boson with quarks of the first two generations. 
Interactions of the Higgs boson with hadronic constituents  
are  then described by  an effective  Lagrangian 
\be
{\cal L} = -C_1 \frac{\alpha_s}{12\pi v} G_{\mu \nu}^{a} G^{a, \mu \nu} H - 
C_2 \frac{m_b}{v} H b \bar b.
\label{eq19}
\ee
The two terms in  Eq.~(\ref{eq19}) refer to 
interactions of Higgs bosons with gluons and $b$-quarks, respectively. 
The first term originates from the $H t \bar t$ interaction and, therefore, is proportional 
to the Higgs-top Yukawa coupling; the second term is proportional to the Higgs-bottom Yukawa 
coupling. 

The two constants  $C_{1,2}$ in Eq.~(\ref{eq19}) are the  
Wilson coefficients of the corresponding operators. 
Their perturbative expansions in the strong  coupling constant -- 
to the order relevant to us -- read (see e.g.~\cite{Davies:2017xsp})
\be
C_1 = -1 +{\cal O}(\alpha_s), 
\;\;\;\; C_2 = 1 + \left ( \frac{\alpha_s}{2\pi} \right )^2 
\left [ \frac{10}{9} - \frac{4}{3}\log \frac{\mu^2}{m_t^2} \right ]
+{\cal O}(\alpha_s^3).
\label{eq13}
\ee

The computation of  NNLO QCD corrections to Higgs boson decay 
to two $b$-quarks reported in Refs.~\cite{Anastasiou:2011qx,DelDuca:2015zqa}
was performed under a tacit  assumption $C_1 =0$ and $C_2 = 1$.  
As we explain below, $C_1 \ne 0$ leads to additional contributions to Higgs decay to $b \bar b$ 
starting at NNLO. In the limit of a small $b$-quark mass, these contributions
scale like $\sim y_b m_b/v \sim m_b^2/v^2$, so they are parametrically indistinguishable from 
terms proportional to $y_b^2$ coming from $C_2$ alone. As a consequence, they
should be included in an NNLO computation.
 However, before discussing this point, we repeat the computation 
of the decay $H \to b \bar b$ reported in Refs.~\cite{Anastasiou:2011qx,DelDuca:2015zqa} 
by setting $C_1 = 0, \; C_2 = 1$.

\subsection{Higgs decay to a $b \bar b$ pair: contribution proportional 
to bottom Yukawa coupling squared}

In this subsection, we  compute NNLO QCD corrections to the decay 
$H \to  b \bar b$ in the approximation  $C_1 = 0$. 
We treat   $b$-quarks 
as massless but with a non-vanishing Yukawa coupling to the Higgs boson. 
The generalization of the  computational method 
described in Ref.~\cite{Caola:2017dug}  to this case 
is straightforward. Since the collinear renormalization of parton distribution 
functions is obviously not needed in this case, 
the computation is simpler 
and the final formulas for the 
NNLO QCD corrections are more compact. There are, 
however, a few subtleties, which we point out in this Section. 

First, as we already mentioned, we work in the   approximation of   massless 
$b$-quarks. This means that the only place where the $b$-quark mass appears is in the 
Yukawa coupling.  We renormalize the Yukawa coupling in the $\overline{\rm MS}$-scheme
at the scale $\mu = m_H$. It is well-known from the computation of the inclusive rate 
that this choice of the renormalization 
scale reduces the magnitude of QCD radiative corrections that 
are very large otherwise~\cite{Braaten:1980yq}. 

Second, integrals of the double-soft eikonal factors are identical to the 
production case and can be re-used in the $H \to b \bar b$ computation. 
Other numerical components of the computation, i.e. integrals of the triple-collinear splitting functions, are  different from the production case 
but they actually become simpler.\footnote{They  are functions of a momentum 
fraction  in the production case 
and just numbers in the decay  case.}

Third, it turns out that the calculation  of {\it the double-collinear 
contributions} is non-trivial for the decay kinematics. 
This is in stark contrast to the computation 
of the NNLO QCD corrections to the production case where the double-collinear contribution 
is among the simplest. The reason for this difference is as follows. The double-collinear 
contributions refer to sectors where collinear singularities  appear 
if, say, the gluon $g_4$ is emitted collinear to the $b$-quark and the gluon 
$g_5$ is emitted collinear to the $\bar b$-anti-quark.  To extract collinear divergences 
in this case, it is convenient to choose  cosines of the relative 
angles between $\vec p_b $ and $\vec p_4$ and between $\vec p_{\bar b}$ and $\vec p_5$ as independent 
kinematic variables. For the decay case, we work in the rest frame of the Higgs boson.
Hence, in contrast to the production case, 
the  directions of $\vec p_b$ and $\vec p_{\bar b}$ are not fixed. It then appears to be 
 non-trivial to use the two angles as independent variables 
and to have 
the phase space properly simplify 
in  soft and collinear limits, while also  
satisfying  the constraint $p_H =  p_4 + p_5 + p_b + p_{\bar b}$.
Nevertheless, this can be done 
and we will present the corresponding formulas in a separate publication. 
Here, we only note that this complexity is a particular feature of the process at hand. 
Since the Born 
process $H\to b \bar b$ involves too few particles, the momentum conservation constraint
makes it difficult to
find a parametrization in terms of the two angles discussed above. For more complicated decay
processes, for example for $Z$ decays to three jets, this issue is not 
present.

The last  point concerns the  contribution of the $b \bar b b \bar b$
 final state to the decay 
rate of the Higgs boson.  This final state  is different from everything that 
we considered so far because we cannot say \textit{a priori} which of the two  $b \bar b$ pairs 
comes from the Higgs vertex and which from the $g^* \to b \bar b$ splitting. Without 
this information, we  cannot separate the phase space into a hard part 
and a radiation part, which is central for the method of Ref.~\cite{Caola:2017dug}. 
To get around this problem, we use the symmetry of the process $H \to b \bar b b \bar b$
with respect to the permutations of the two  $b$-quarks and the two $\bar b$-anti-quarks 
and split the matrix element into a part that is equivalent to the singlet component
$H \to b \bar b + q \bar q$, $q\ne b$,  
and an identical quark interference contribution. In each of the interference contributions, there is 
either  a  quark line or an anti-quark 
line that always originates from the Higgs decay  vertex. We assign this line to belong 
to the hard phase space. 
The remaining lines  can originate either 
from  the Higgs decay vertex or  from the $g^* \to b \bar b$ splitting.
Which line   belongs to the hard phase space and which one to the radiative phase space is 
a matter of choice at this point. The interference terms only contain a purely triple-collinear
singularity. It 
 corresponds to the interference term in the non-singlet triple-collinear 
splitting function \cite{Catani:1999ss}
and can be easily extracted and integrated numerically. 
 
We continue by presenting some numerical results of the calculation. Again, 
our goal in this Section is not to discuss phenomenology of the Higgs boson 
decay to a $b \bar b$ pair but to show that our method 
is capable of producing high-precision results. 

The numerical computation yields the following result 
for the decay rate of the Higgs boson to a $b \bar b$ pair 
\be
\Gamma(H \to b \bar b) = \Gamma_{\rm LO} 
\left [ 1 + \left ( \frac{\alpha_s}{2\pi} \right ) 11.3333(16)
+ \left ( \frac{\alpha_s}{2\pi} \right )^2 116.68(8) + ... \right ],
\label{eq12}
\ee
where $\Gamma_{\rm LO} = 3 y_b^2  m_H/(16 \pi) = 3 m_b^2(m_H) m_H/(8\pi v^2)$. The value of the Yukawa 
coupling constant has already been discussed in the previous Section. 
The renormalization scale for the strong coupling constant is set to the mass 
of the Higgs boson.

It is instructive to compare Eq.~(\ref{eq12})
with the results of an analytic computation~\cite{baikov}. 
The analytically-known 
two-loop coefficient evaluates to  $116.59...$ which is in better 
than per mill  agreement with the result of the numerical 
computation shown in Eq.~(\ref{eq12}).

It is also interesting to compute jet rates in $H \to b \bar b$ decay since 
such,  more exclusive, calculations  provide 
a stronger test of the numerical stability  of the method. 
Similar to Ref.~\cite{Anastasiou:2011qx},  
we use the JADE clustering algorithm with 
$y_{\rm cut} = 10^{-2}$ to define jets.\footnote{Following Ref.~\cite{Anastasiou:2011qx},
we define the JADE distance as $y_{ij} = (p_i+p_j)^2$.} We obtain 
\be
\begin{split} 
& \Gamma_{2j}  = \Gamma_{\rm LO}  \left [ 
1 - 27.176(3) \lp\frac{\alpha_s}{2\pi}\rp  - 1240.78(21) \left ( \frac{\alpha_s}{2\pi} \right)^2 
+\mathcal O(\alpha_s^3)
\right],
\\
& \Gamma_{3j}  = \Gamma_{\rm LO}  \left [ 
38.509(3) \left ( \frac{\alpha_s}{2\pi} \right )   + 980.61(10)  \left ( 
\frac{\alpha_s}{2\pi}  \right )^2  +\mathcal O(\alpha_s^3)\right ],
\\
& \Gamma_{4j} =    376.784(8)  \; 
\left ( \frac{\alpha_s}{2\pi} \right )^2 \Gamma_{\rm LO} + \mathcal O(\alpha_s^3).
\end{split}
\label{eq14}
\ee
The sum of the exclusive jet rates in Eq.~(\ref{eq14}) gives the total decay rate; computing 
this sum, we obtain 
\be
\Gamma(H \to b \bar b) = \Gamma_{2j} + \Gamma_{3j} + \Gamma_{4j} 
= 
\Gamma_{\rm LO} 
 \left [ 1 + \left ( \frac{\alpha_s}{2 \pi} \right ) 11.3334(41)
 + \left ( \frac{\alpha_s}{2 \pi} \right )^2 116.62(23) + ... \right ].
\label{eq14a} 
\ee
Comparing the inclusive computation shown in Eq.~(\ref{eq12}) with the 
sum of exclusive jet rates in Eq.~(\ref{eq14a}), we find perfect 
agreement although the integration error is 
somewhat larger in the  latter case. 

\subsection{Additional contributions to Higgs decay proportional to top Yukawa coupling}
\label{subsect:Hbb}

We mentioned above that a non-vanishing Wilson coefficient $C_1$ gives rise to additional contributions to $H \to b \bar{b}$ decays starting at NNLO in QCD, which were not considered in previous 
fully-differential calculations~\cite{DelDuca:2015zqa,Anastasiou:2011qx}. We describe these contributions in more detail in this subsection.
These 
contributions are of the interference type: an amplitude where the Higgs 
boson decays to two (real or virtual) gluons that later turn into  bottom quarks 
interferes with an amplitude where the Higgs boson decays directly to bottom quarks 
and gluons.   

\begin{figure}[t]
\centering
\includegraphics[width=1.\textwidth]{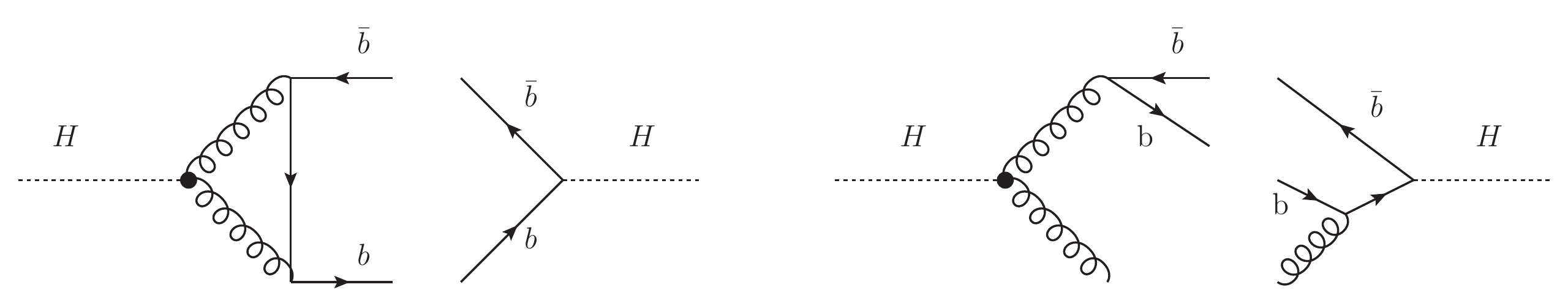}
\caption{Illustrative interference diagrams that contribute to the $H\to b \bar b$ decay rate
for $C_1\ne 0$. See text for details.}\label{Hintdiags}
\end{figure}

Some of these contributions  are shown 
in Fig.~\ref{Hintdiags}.  
They are proportional to the product of 
two Wilson coefficients $C_1 C_2$ 
and, therefore, 
to the {\it first power} of the $b$-quark Yukawa coupling, 
at variance  with   contributions to $H \to b \bar b$ 
decay considered in  the previous subsection.
However, 
angular momentum conservation implies that diagrams in 
Fig.~\ref{Hintdiags} can interfere only  if a  helicity flip 
occurs on one of the $b$-quark lines; effectively, this helicity 
flip and  the Wilson coefficient $C_1$   provide another factor $m_b/v$,  
making the overall scaling of these 
interference contributions with 
the $b$-quark mass identical to 
what we have seen  in the previous subsection. 

These contributions are soft and collinear 
finite for $m_b \ne 0$. Indeed, taking the real emission contribution 
as an example, it is easy to see that the collinear singularity associated 
with the splitting $g^* \to b \bar b$ is regulated because the gluon 
invariant mass should exceed $2m_b$.  Similar 
considerations 
ensure that the virtual diagram shown in Fig.~\ref{Hintdiags} has no soft and collinear 
divergences for finite  $m_b$ as well. 

However, since the  calculation  in the previous subsection  was performed with  massless 
$b$-quarks, we would like to compute the diagrams 
shown in Fig.~\ref{Hintdiags} in the same approximation.  Unfortunately, doing so leads to problems. 
Indeed, if we factor  out one power of $m_b$ caused by 
the helicity flip, the reduced  matrix 
element   has peculiar 
soft and collinear limits 
in the $m_b = 0$ approximation, 
that are typically not present in QCD amplitudes at leading power. 
For example,
it develops  a logarithmic singularity 
when a \emph{single} $b$-quark becomes  soft. 

The validity of the massless approximation assumes that 
the logarithmic dependence on the
$b$-quark mass cancels out in infra-red safe quantities. 
It is easy to see, however,  that this
cancellation does not take place for the interference contributions, 
and it is not possible to give a proper inclusive definition of 
this process in the massless approximation. Indeed, the logarithmic mass dependence cancels
between the diagrams in Fig.~\ref{Hintdiags} \emph{and} similar diagrams 
with a $b$-mediated Higgs decay into gluons. One could try to circumvent this problem
by regulating 
the collinear singularity related to $g^*\to b\bar b$ with a  
flavored jet algorithm, e.g. the one in Ref.~\cite{Banfi:2006hf}. 
This would trade the logarithmic dependence on  the $b$-quark mass
for  a logarithmic sensitivity to  a jet radius $R$. 
However, even this does not solve the problem completely 
as the single-soft quark singularity is {\it not} regulated 
by the jet algorithm of Ref.~\cite{Banfi:2006hf}. 

It is clear that a proper description of  the 
interference contributions requires a computation  with 
fully massive $b$-quarks. In its absence, 
we estimate the order of magnitude of these effects by simply imposing 
restrictions on the phase space of the $b$-quarks and gluons
that reproduce  the leading logarithmic terms. 
We find that these contributions 
may change
the NNLO corrections to the inclusive Higgs 
decay rate shown in Eq.~(\ref{eq12}) by up to ${\cal O}(30\%)$.
Using the same setup in the  fiducial region
that will be discussed in the next Section, we find 
that this interference contribution is  somewhat reduced. 
Since their impact on the decay rate appears to be limited, 
we will  omit these terms  from the 
phenomenological 
analysis in the next  Section but we stress that it is 
important to understand them better. As we explained,  this will require a
  fully-differential computation of the Higgs 
decay to massive bottom pairs at NNLO.  We leave this for 
future investigations.

\section{ The physical process} 
\label{sect:fullpheno}

We are now in position to discuss the physical 
process $pp \to  W (l \nu) H ( b \bar b)$,  
including  QCD corrections to both production and decay. 
Given the results of the preceding Sections, it is straightforward 
to do so. The only subtlety is how to treat the Higgs boson decay width 
that appears in the cross section in the narrow width approximation. We write
\be
{\rm d} \sigma_{WH(bb)} = {\rm d} \sigma_{WH} \times \frac{{\rm d} \Gamma_{bb}}{\Gamma_H} 
= {\rm Br}(H \to b \bar b)  \times 
{\rm d} \sigma_{WH} \times \frac{{\rm d} \Gamma_{bb}}{\Gamma_{bb}}. 
\label{eq18}
\ee
We note that in the approximation of massless $b$-quarks, 
the Higgs boson decay rate to a $b \bar b$ pair 
and {\it therefore the Higgs branching ratio to a $b \bar b $-pair } subtly depends on the 
definition of a $b$-quark. However, the effect on the total decay rate is relatively small, 
as discussed 
in the previous Section, and we set $C_1=0$ for the phenomenological studies in this
paper. We use 
 ${\rm Br}(H \to b \bar b) = 0.5824$~\cite{deFlorian:2016spz} as a fixed quantity, not subject
to an $\alpha_s$ expansion. 

To define an expansion of Eq.~(\ref{eq18}) in $\alpha_s$, we follow 
Ref.~\cite{Ferrera:2017zex},
write the production cross section and the decay width to $b \bar b$ 
as an expansion in $\alpha_s$
\be
{\rm d} \sigma_{WH} = \sum_{i=0}^{\infty} {\rm d} \sigma_{WH}^{(i)},
\;\;\;\;\;\;
{\rm d} \Gamma_{b \bar b } = \sum_{i=0}^{\infty} {\rm d} \Gamma_{b \bar b}^{(i)},
\ee
and introduce 
\be
{\rm d} \gamma^{(i)} = \frac{\sum \limits_{j=0}^{i} {\rm d} \Gamma_{b \bar b}^{(j)}}{\sum \limits_{j=0}^{i} \Gamma_{b \bar b}^{(j)}}.
\ee
Note that ${\int } {\rm d} \gamma^{(i)} = 1$, 
provided that the integration goes over the unrestricted phase space.  

Using this notation, we define the physical cross sections computed 
through  different orders in QCD 
perturbation theory 
\be
\begin{split} 
& {\rm d} \sigma_{WH(b \bar b)}^{\rm LO} = {\rm Br}(H \to b \bar b) \;
{\rm d} \sigma^{(0)} \; {\rm d} \gamma^{(0)},
\;
\\
& {\rm d} \sigma_{WH(b \bar b)}^{\rm NLO} = {\rm Br}(H \to b \bar b) \left [ 
{\rm d} \sigma^{(0)} \; {\rm d} \gamma^{(1)} 
+ {\rm d} \sigma^{(1)} \; {\rm d} \gamma^{(0)}  \right ],
\\
& {\rm d} \sigma_{WH(b \bar b)}^{\rm NNLO} = {\rm Br}(H \to b \bar b) \left [ 
{\rm d} \sigma^{(0)} \; {\rm d} \gamma^{(2)} 
+ {\rm d} \sigma^{(1)} \; {\rm d} \gamma^{(1)}  
+ {\rm d} \sigma^{(2)} \; {\rm d} \gamma^{(0)}  
\right ].
\label{eq21}
\end{split}
\ee
In addition, for comparison with the previous computations of Refs.~\cite{Ferrera:2013yga,Campbell:2016jau}, 
it is convenient to 
introduce an approximate NNLO cross section that includes  NNLO corrections 
to the production process but only NLO corrections to the decay. It reads 
\be
{\rm d} \sigma_{WH(b \bar b)}^{\rm NNLO,\rm approx} = {\rm Br}(H \to b \bar b) \left [ 
{\rm d} \sigma^{(0)} \; {\rm d} \gamma^{(1)} 
+ {\rm d} \sigma^{(1)} \; {\rm d} \gamma^{(0)}  
+ {\rm d} \sigma^{(2)} \; {\rm d} \gamma^{(0)}  
\right ].
\label{eq22}
\ee

We are now in a position to discuss the results of the computation.  To define 
the $W(l\nu)H(b \bar b)$ final state, we reconstruct $b$-jets using the infra-red
safe flavor-$k_t$ jet algorithm~\cite{Banfi:2006hf}\footnote{We are grateful to G.~Salam
for providing us with his private implementation of the algorithm~\cite{Banfi:2006hf} within the
\texttt{FastJet} framework~\cite{Cacciari:2011ma}.} with $\Delta R = 0.5$
and require that an event should contain at least one $b$-jet  and one 
$\bar b$-jet  with
\be
|\eta_{j_b}| < 2.5,\;\;\;\; p_{\perp,j_b} > 25~{\rm GeV}.
\ee
An identified light (non-$b$) jet is required to have a transverse 
momentum $p_\perp > 25~{\rm GeV}$
as well but no  pseudo-rapidity cut is applied in this case. 
In addition, 
we impose the following cuts on the pseudorapidity and transverse momentum of 
the charged lepton
\be
|\eta_{l}| < 2.5,\;\;\;\;p_{\perp,l} > 15~{\rm GeV}.
\ee

Finally, following the experimental analyses, 
we may impose an additional requirement that the vector boson  has a 
transverse momentum $p_{\perp,W} > 150~{\rm GeV}$.   We use parton distribution 
functions NNPDF3.0 as in Section~\ref{sect:WHprod}.  However, at variance 
with the 
calculation reported there, here
we employ LO, NLO and NNLO PDFs to compute LO, NLO and NNLO 
cross sections, respectively.

We begin by presenting the fiducial volume cross sections for the process 
$pp \to W(l \nu) H(b \bar b)$ at the 13 TeV LHC, at various orders in perturbative QCD.
The $W^+$ case has already been studied in Ref.~\cite{Ferrera:2017zex}. 
For this reason here we focus on the 
$W^-$ case. Without the cut on $p_{\perp,W}$, we find 
\be
\begin{split}
& \sigma^{{\rm LO}}_{{\rm fid},W^-} = 15.50_{-0.56}^{+0.44}~\rm{fb},
~~~~
\sigma^{{\rm NLO}}_{{\rm fid},W^-} = 16.13_{+0.20}^{-0.09}~\rm{fb},\\
& \sigma^{{\rm NNLO}}_{{\rm fid},W^-} = 15.20_{+0.11}^{-0.08}~\rm{fb},
~~~~
\sigma^{{\rm NNLO,approx}}_{{\rm fid},W^-} = 16.56_{+0.16}^{-0.11}~\rm{fb}.
\end{split} 
\label{eq23}
\ee
Imposing the cut on the transverse momentum 
of the $W$ boson $p_{\perp,W} > 150~{\rm GeV}$, we obtain
\be
\begin{split}
& \sigma^{{\rm LO}}_{{\rm fid},W^-} = 2.027_{+0.006}^{-0.013}~\rm{fb},
~~~~
\sigma^{{\rm NLO}}_{{\rm fid},W^-} = 2.381_{+0.055}^{-0.041}~\rm{fb},\\
& \sigma^{{\rm NNLO}}_{{\rm fid},W^-} = 2.357_{+0.018}^{-0.026}~\rm{fb},
~~~~
\sigma^{{\rm NNLO,approx}}_{{\rm fid},W^-} = 2.516_{+0.025}^{-0.030}~\rm{fb}.\\
\label{eq24}
\end{split}
\ee
For the cross sections in Eqs.~(\ref{eq23},\ref{eq24}) 
the central value corresponds to the factorization
and renormalization scales {\it in the production process} set to 
the invariant mass of the $WH$ system. The uncertainties are obtained
by changing simultaneously the renormalization and factorization scales in the
production process by a factor of two, $\mu_R=\mu_F=\{1/2,1,2\}\times m_{WH}$.
As we said already, this is most likely an underestimate of the total theory
uncertainty, but this issue has already been discussed at length in the
literature (see e.g.~\cite{deFlorian:2016spz}) and it is not the point of our 
study. Consequently, we do not comment on it any further and, in what follows,
we only show distributions for the central scale choice. 
For the decay process, we always use the scale $\mu = m_H$. 

The results for the fiducial cross sections Eqs.~(\ref{eq23},\ref{eq24}) show that NLO QCD effects 
are larger if a transverse momentum cut is imposed on the $W$ boson.
This is expected since the $W$ boson can evade this cut by recoiling against 
additional radiation which appears at NLO.
The approximate NNLO results, which include NNLO corrections to the production process only, show a similar effect: this cross section is about 3\% higher than the NLO cross section without the $p_{\perp,W}$ cut, but about 6\% higher when this cut is imposed.
Including Higgs decay at NNLO decreases the approximate cross section
by about 9\% without the $p_{\perp,W}$ cut, and 7\% in the presence of this cut. Therefore
there are cancellations between corrections to the production and decay sub-processes,
that make the size of the full NNLO QCD corrections quite sensitive to the value of the
$p_{\perp,W}$ cut.

\begin{figure}[t]
\centering
\includegraphics[width=0.485\textwidth]{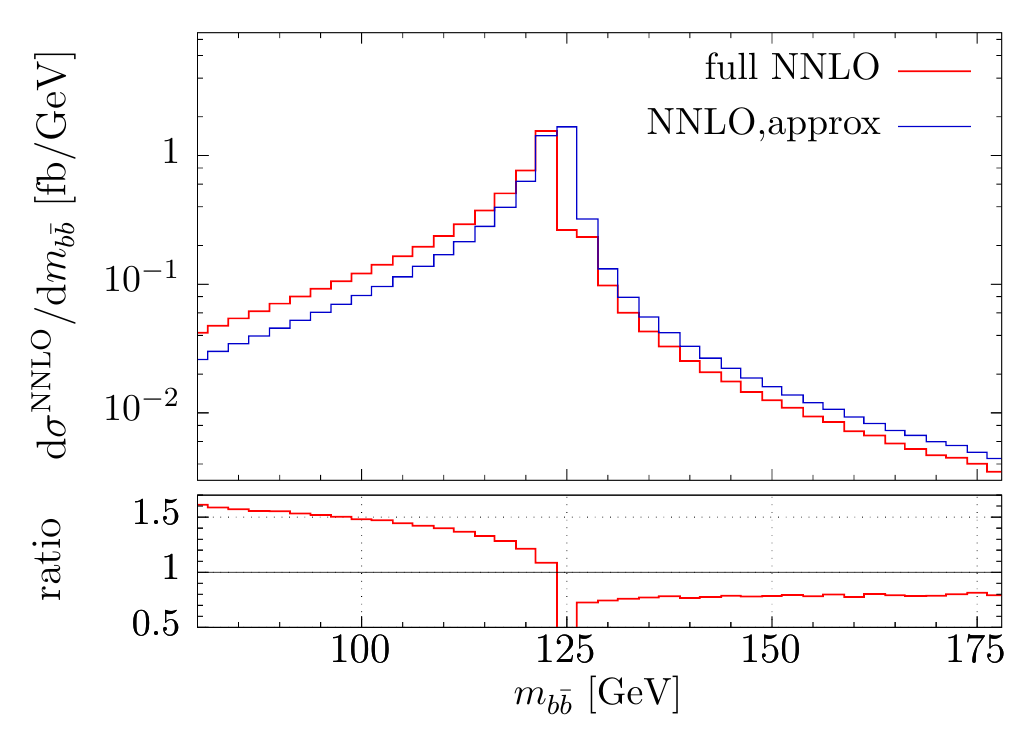}
~~
\includegraphics[width=0.485\textwidth]{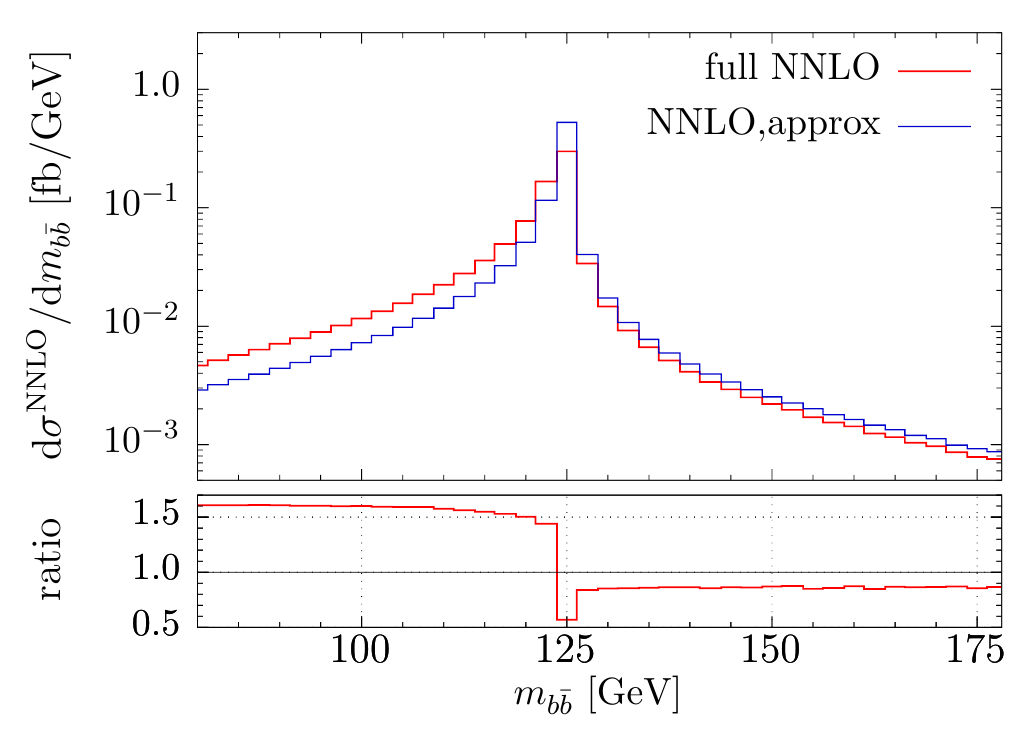}
\caption{The invariant mass of a 
$b$-jet and a $\bar b$-jet  that best approximates 
the Higgs boson mass. 
Left pane -- without the $p_\perp^{W}$ cut, right pane -- with 
the $p_{\perp}^{W} > 150~{\rm GeV}$ cut. Lower panes -- ratio of full NNLO 
to approximate NNLO. The renormalization and factorization scales are set 
to $\mu_R=\mu_F = M_{WH}$ for the production process and to $\mu_R = m_H$ for the decay
process.  See text for further details.} 
\label{mHbb}
\end{figure}

We now turn to differential distributions. 
We begin by identifying the $b \bar b$ system comprised 
of a $b$-jet and a $\bar b$-jet whose invariant mass 
best approximates the mass of the Higgs boson, and
consider 
the invariant mass $m_{b \bar b}$ distribution  of this 
 $b \bar b$ system. Since we work 
in the narrow width approximation, 
at leading order this distribution is described by a 
delta-function  $\delta(m_{b \bar b}^2 - m_H^2)$. 
At next-to-leading order, this situation changes: 
a gluon emitted in the Higgs boson decay can decrease the invariant mass of the $b \bar b$ system 
while a gluon emitted in the production process can increase it.  Hence, the $m_{b \bar b}$ distribution has tails both above and  below 
$m_{b\bar b}= m_H$ that start 
to appear if the next-to-leading order correction 
to either  production or decay  is included in the computation.  In 
Fig.~\ref{mHbb} we compare predictions for this observable  obtained using 
full and approximate NNLO computations, defined in Eqs.~(\ref{eq21},\ref{eq22}), 
respectively. We study this observable both without (left) and with (right) 
the cut on the $W$ boson transverse momentum $p_{\perp,W}>150~{\rm GeV}$.
It is seen from Fig.~\ref{mHbb}  that the application of this 
cut affects the  shape of $m_{b \bar b}$ distribution in a minor way. 
For example, in  both cases, 
 full NNLO results 
deplete the distribution at $m_{b \bar b} > m_H$ and enhance the distribution
at $m_{b \bar b} < m_H$  relative  to  approximate NNLO predictions.
Since the full  NNLO  provides a better 
description of the  
radiation in the decay, compared to the approximate NNLO, 
and since radiation in the decay predominantly reduces $m_{b \bar b}$, this re-shaping is not unexpected. However,  
the magnitude of this ${\cal O}(\alpha_s^2)$  effect --  
${\cal O}(60 \%)$ 
correction at $m_{b \bar  b} \sim 80~{\rm GeV}$ and 
${\cal O}(-15\%)$ at $m_{b \bar b} > m_H$ --   is 
somewhat  surprising. We note that similarly large corrections
have also been observed in Ref.~\cite{Ferrera:2017zex}.

\begin{figure}[t]
\centering
\includegraphics[width=0.485\textwidth]{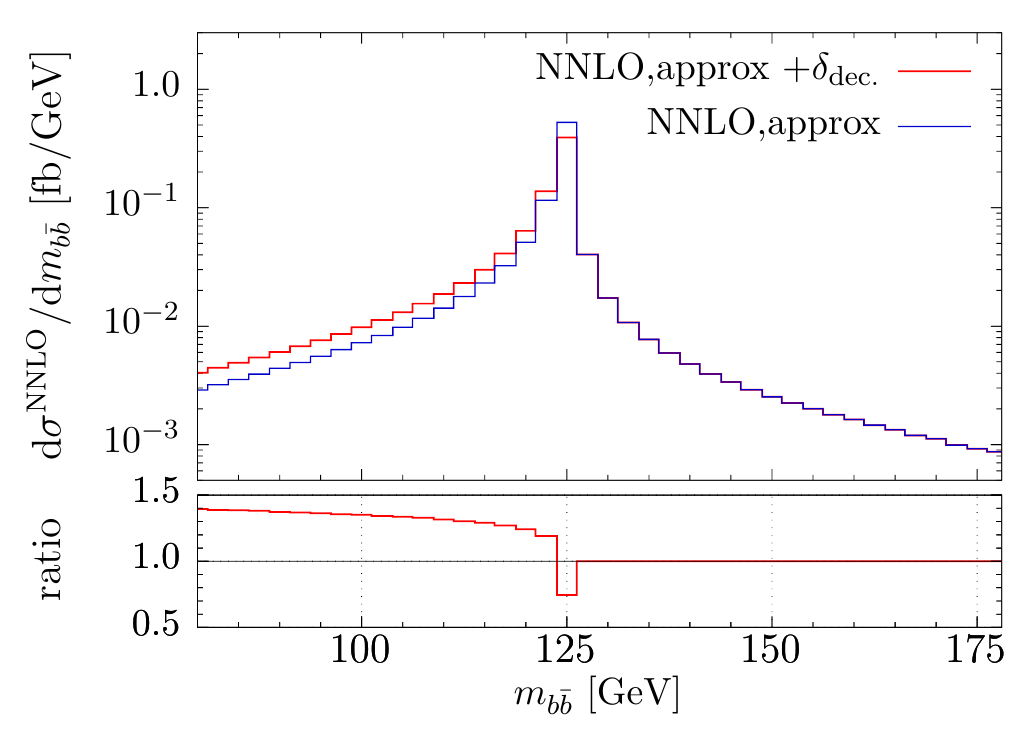}
~~
\includegraphics[width=0.485\textwidth]{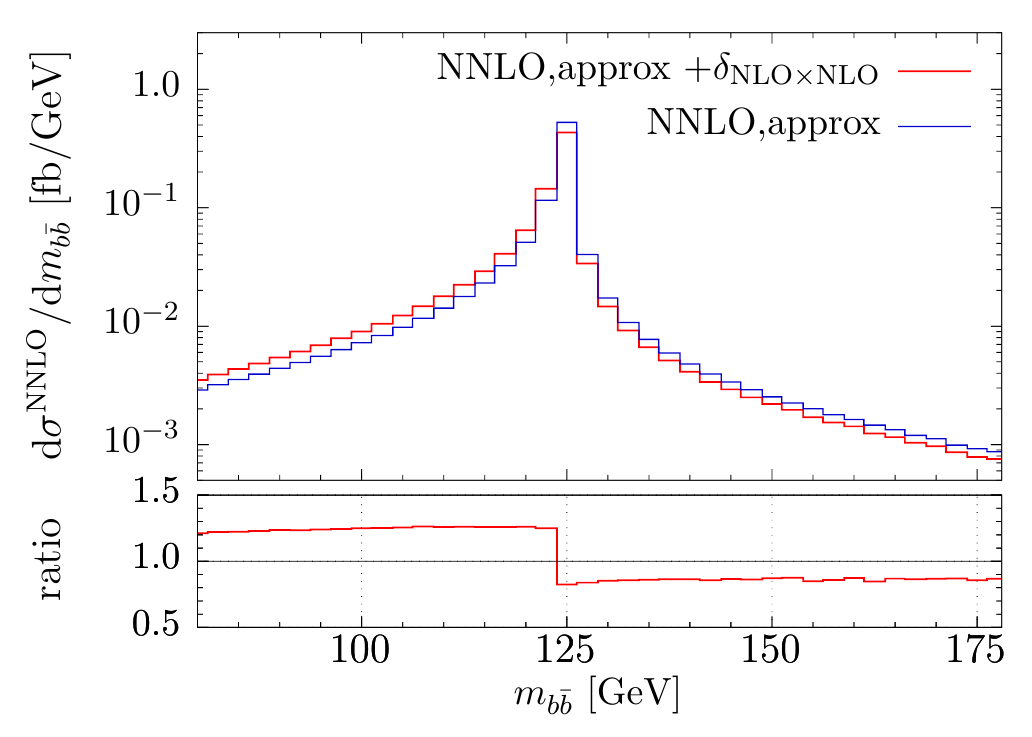}
\caption{The invariant mass of a $b$-jet and a $\bar b$-jet
that  best approximates  the Higgs boson mass. The 
$p_{\perp}^{W} > 150~{\rm GeV}$ cut is applied.  Left pane: only NNLO corrections 
to decay are included. 
 Right pane: NLO corrections to the production and NLO corrections 
to the decay are included. 
Lower panes -- ratio 
to approximate NNLO. See text for further details. 
} 
\label{mHbba}
\end{figure}

To understand what causes these large effects, we 
split the difference between approximate 
NNLO and full NNLO into two terms -- 
NNLO radiation in the decay (${\rm NNLO}_{\rm dec}$) 
and  NLO radiation in the production followed by 
the NLO radiation in the decay (${\rm NLO}_{\rm prod} \times {\rm NLO}_{\rm dec}$).
We define
\be
\begin{split}
  \delta_{\rm dec.} =&  {\rm Br}(H \to b \bar b) \;
        {\rm d} \sigma^{(0)} \left( {\rm d} \gamma^{(2)}-{\rm d} \gamma^{(1)} \right), \\
\delta_{{\rm NLO}\times{\rm NLO}} =&  {\rm Br}(H \to b \bar b) \;
        {\rm d} \sigma^{(1)} \left( {\rm d} \gamma^{(1)}-{\rm d} \gamma^{(0)} \right), \\
\end{split}
\ee
such that  ${\rm d} \sigma_{WH(b \bar b)}^{\rm NNLO,\rm approx} + \delta_{\rm dec.}+\delta_{{\rm NLO}\times{\rm NLO}} = {\rm d} \sigma_{WH(b \bar b)}^{\rm NNLO}$.
We display the two distributions 
in~Fig.~\ref{mHbba}.   As we said already, the radiation in the decay does 
not populate the $m_{b \bar b}$ region to the right of $m_H$,   so that 
the $\mathcal O(-15\%)$ correction at such values of the $b \bar b$ invariant 
mass 
comes exclusively from the ${\rm NLO}_{\rm prod} \times {\rm NLO}_{\rm dec}$ contribution. 
On the other hand, for $m_{b \bar b} < m_H$ the NNLO corrections to the decay play a dominant role, increasing the
distribution by about 40\%, as compared to the $\mathcal{O}(20\%)$ increase from ${\rm NLO}_{\rm prod} \times {\rm NLO}_{\rm dec}$.

\begin{figure}[t]
\centering
\includegraphics[width=0.485\textwidth]{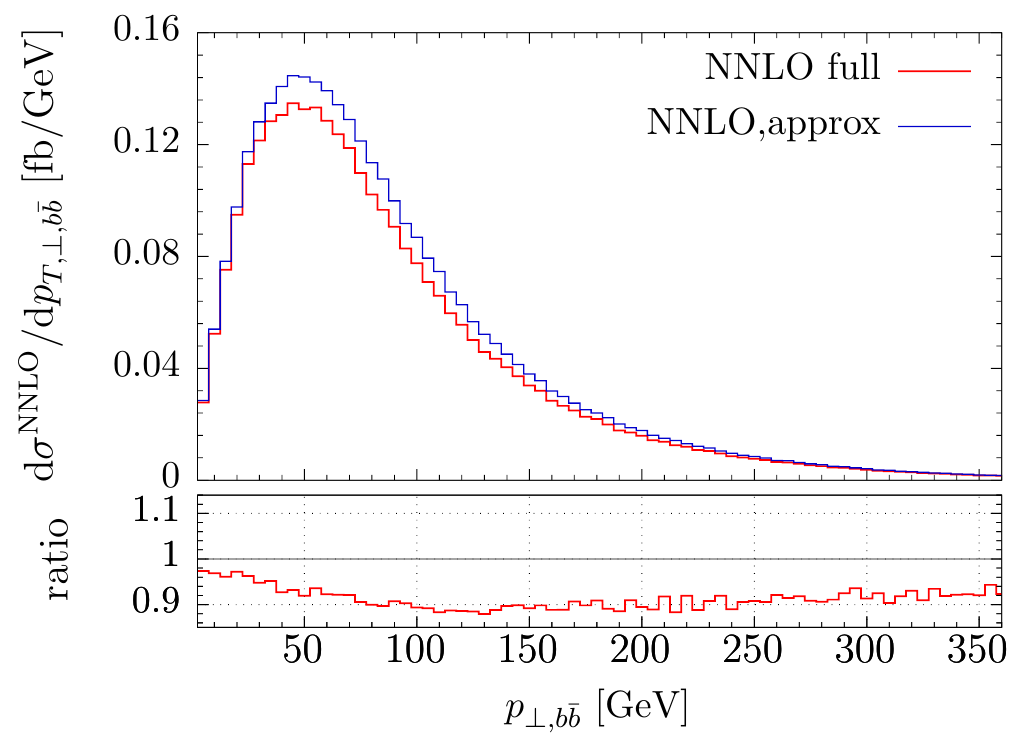}
~~
\includegraphics[width=0.485\textwidth]{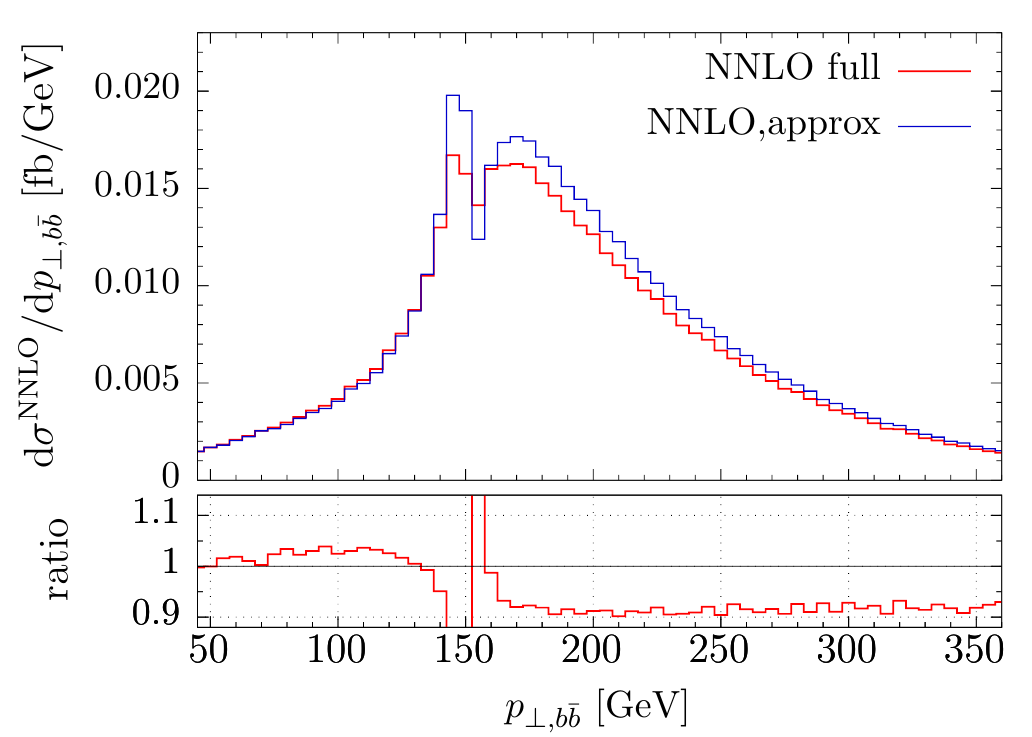}
\caption{
Same as Fig.~\ref{mHbb}, but for the transverse momentum of the $b\bar b$
system that is used to reconstruct the Higgs boson. See text for further
details. 
} 
\label{pTbb}
\end{figure}

Next, we consider the transverse momentum  of the $b \bar b$ system whose 
invariant mass provides the best approximation to the Higgs boson mass.  The 
NNLO and approximate NNLO distributions for this observable 
are compared  in Fig.~\ref{pTbb}; 
the cut $p_{\perp,W} > 150~{\rm GeV}$  
is applied to events displayed in the right pane. 
It follows from Fig.~\ref{pTbb} 
that the cut on the $W$ boson transverse momentum 
re-shapes the distribution, pushing 
its maximum to larger values.
Again, this is easily understood by observing that the
$p_{\perp,W}$ cut implies the requirement $p_{\perp, b \bar b} > 150~\rm{GeV}$ at LO.
In addition, if the cut 
on the $W$ transverse momentum is applied, 
both the full and the approximate NNLO calculations  
develop a Sudakov shoulder  below 
$p_{\perp,b \bar b} = p_{\perp,W}^{\rm cut} = 150~{\rm GeV}$.
We note that this feature is somewhat less prominent in the
full NNLO distribution. 

\begin{figure}[t]
\centering
\includegraphics[width=0.485\textwidth]{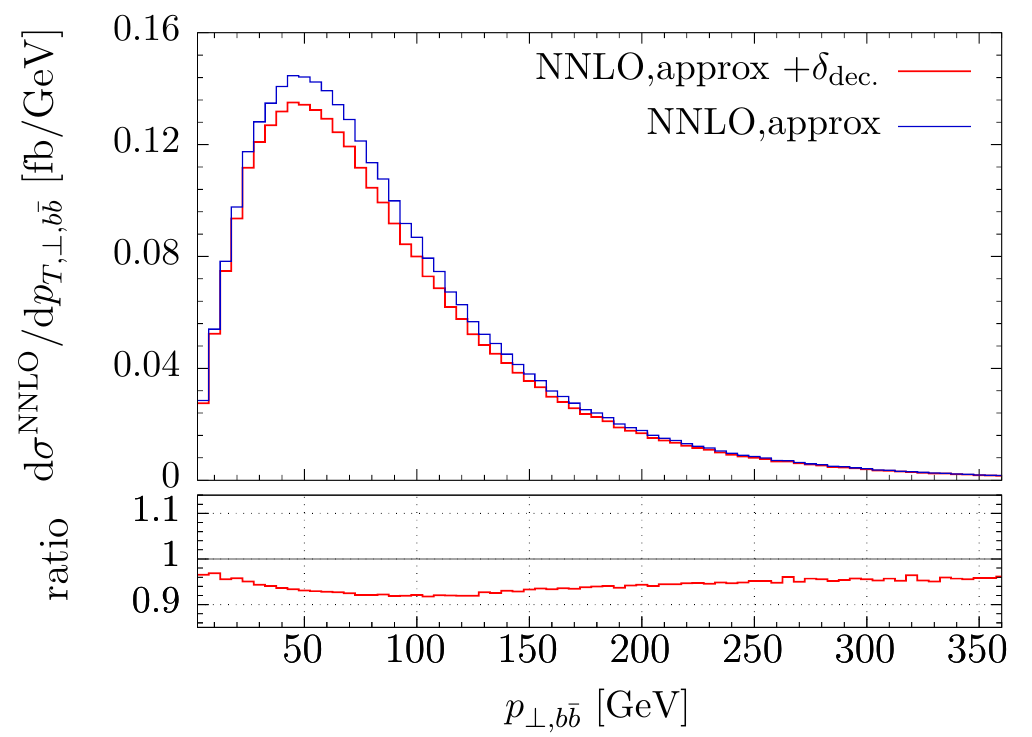}
~~
\includegraphics[width=0.485\textwidth]{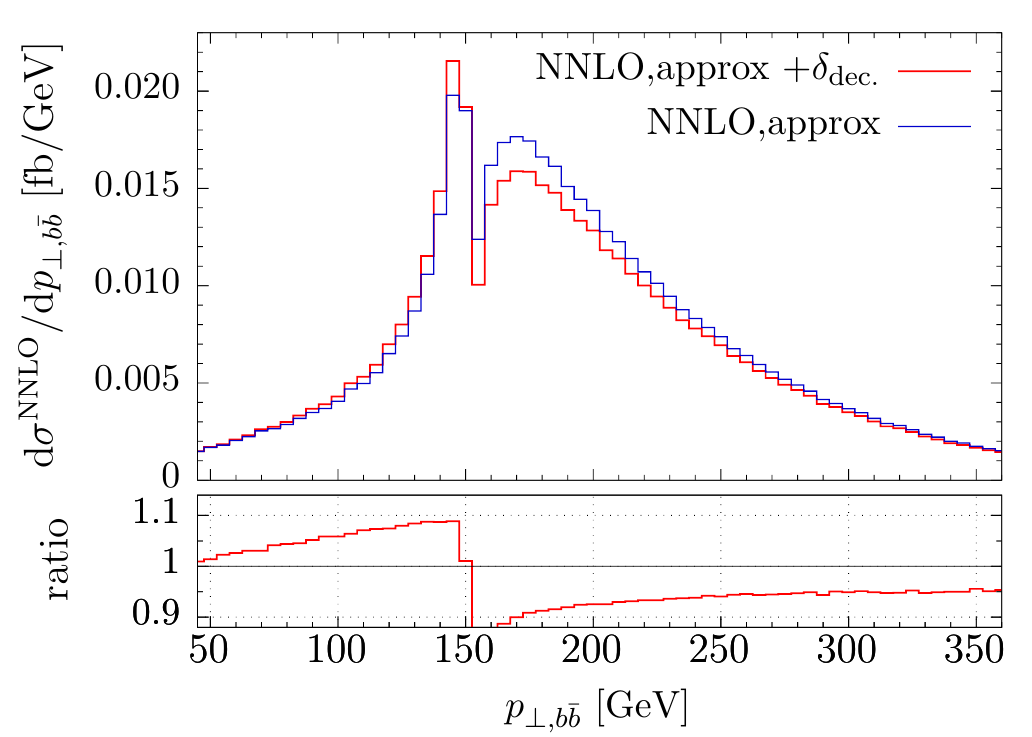}
\includegraphics[width=0.485\textwidth]{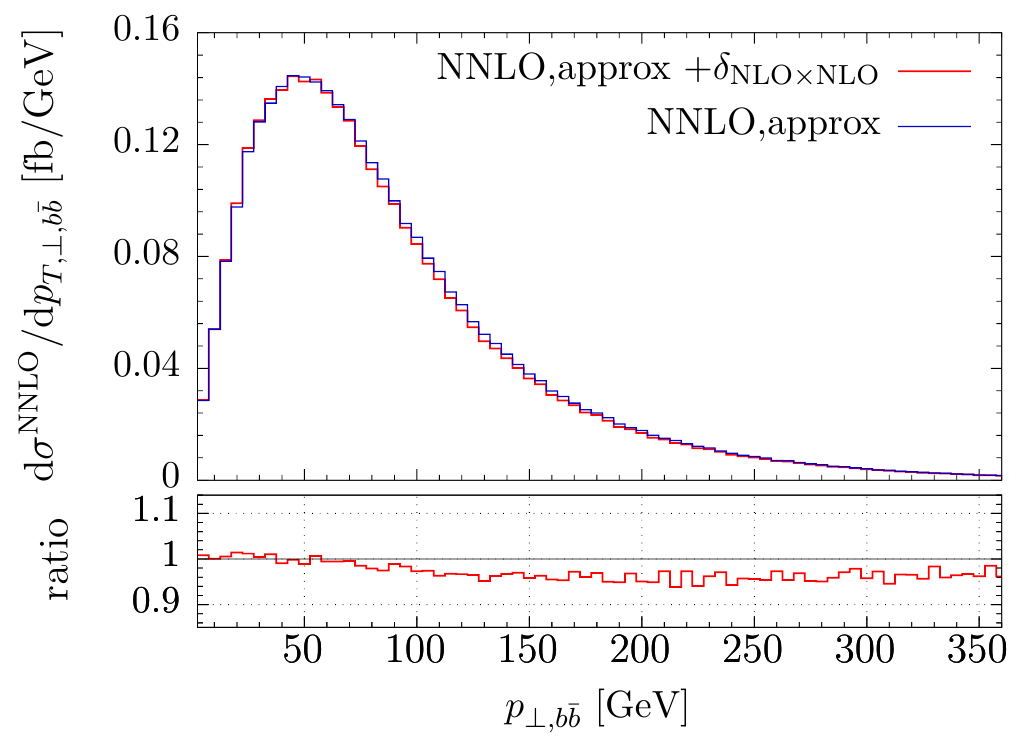}
~~
\includegraphics[width=0.485\textwidth]{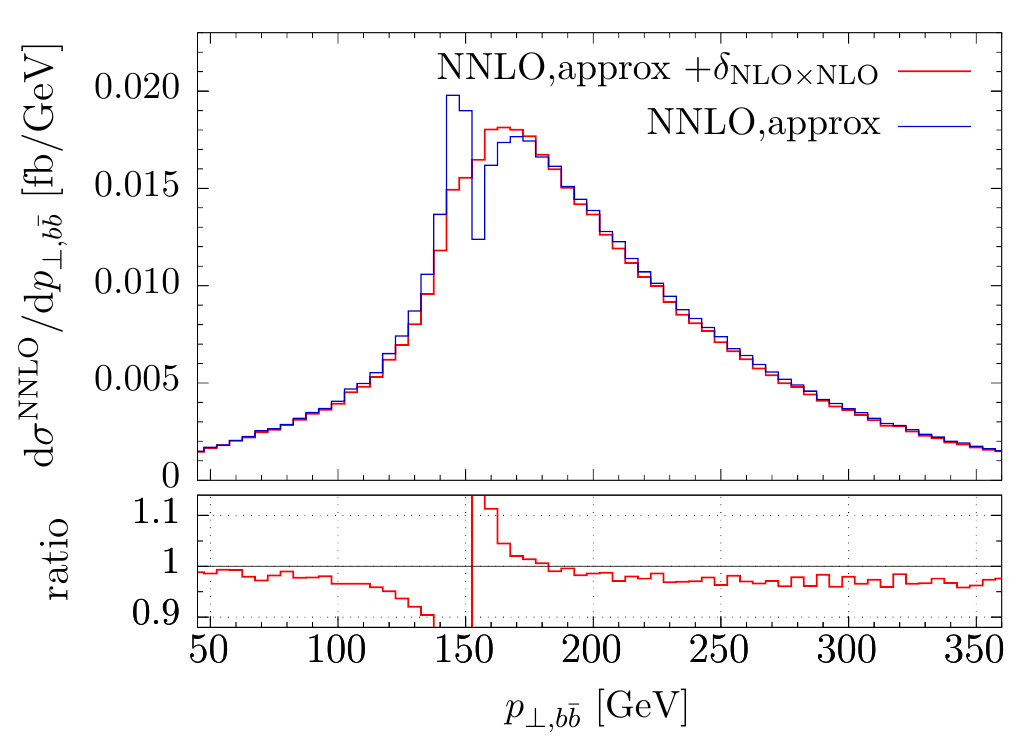}
\caption{The different contributions to 
the distribution of the sum of transverse momenta of the 
$b$- and $\bar b$-jets that are used to reconstruct the Higgs boson.
Left pane -- without the $p_\perp^{W}$ cut, right pane -- with 
the $p_{\perp}^{W} > 150~{\rm GeV}$ cut.  Upper row: only NNLO corrections
to decay are included. Lower row: NLO corrections to the production and
NLO corrections to the decay are included. 
See text for further details. } 
\label{pTbb_ab}
\end{figure}

To understand the relative impact of different contributions, we again split the 
full NNLO into two different 
parts, $\delta_{\rm dec.}$ and $\delta_{{\rm NLO}\times{\rm NLO}}$, 
and display them  separately in Fig.~\ref{pTbb_ab}.
For values of $p_{\perp,b \bar b}$ 
larger than $p_{\perp,W}^{\rm cut}$, the approximate NNLO is larger than the 
full NNLO by about ${\cal O}(5\%-10\%)$, independent of whether or not the 
cut on the $W$ boson transverse momentum  is applied, due to the 
corrections from both  ${\rm NLO}_{\rm prod} \times {\rm NLO}_{\rm dec}$
and the NNLO decay. When the $p_{\perp,W}$ cut is imposed, the slight increase at low values of
 $p_{\perp,b \bar b}$ is the result of a cancellation between the 
somewhat larger contributions from the NNLO decay and the 
${\rm NLO}_{\rm prod}\times{\rm NLO}_{\rm dec}$.
We also note that the ${\rm NLO}_{\rm prod}\times{\rm NLO}_{\rm dec}$ contribution smears
the Sudakov shoulder.

\begin{figure}[t]
\centering
\includegraphics[width=0.485\textwidth]{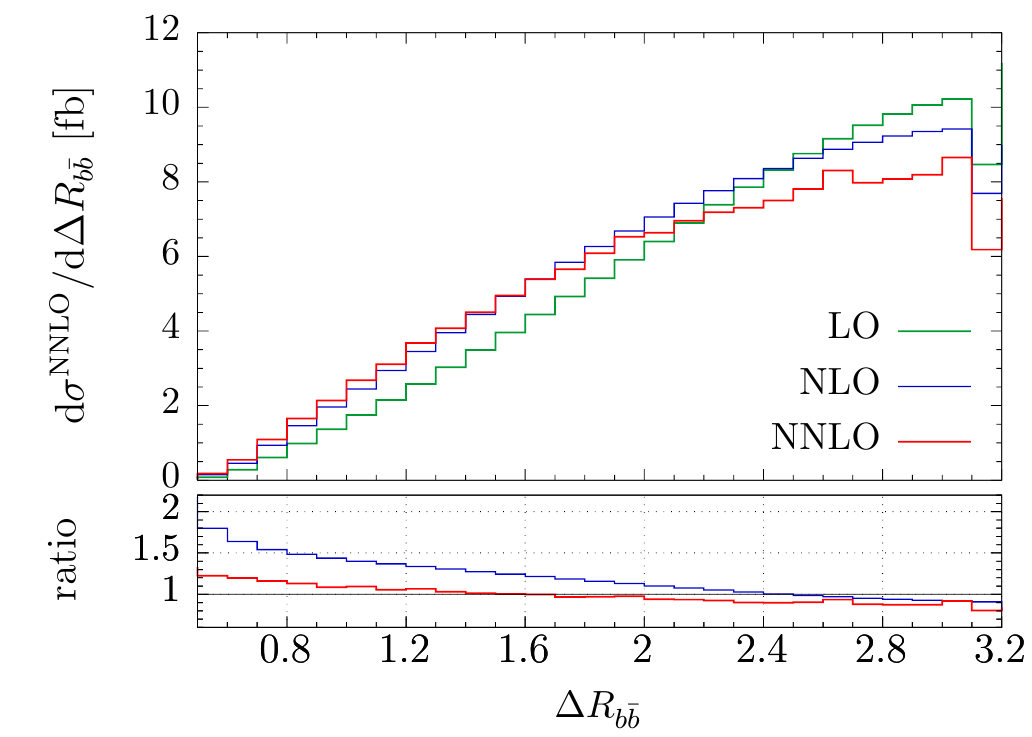}
~~
\includegraphics[width=0.485\textwidth]{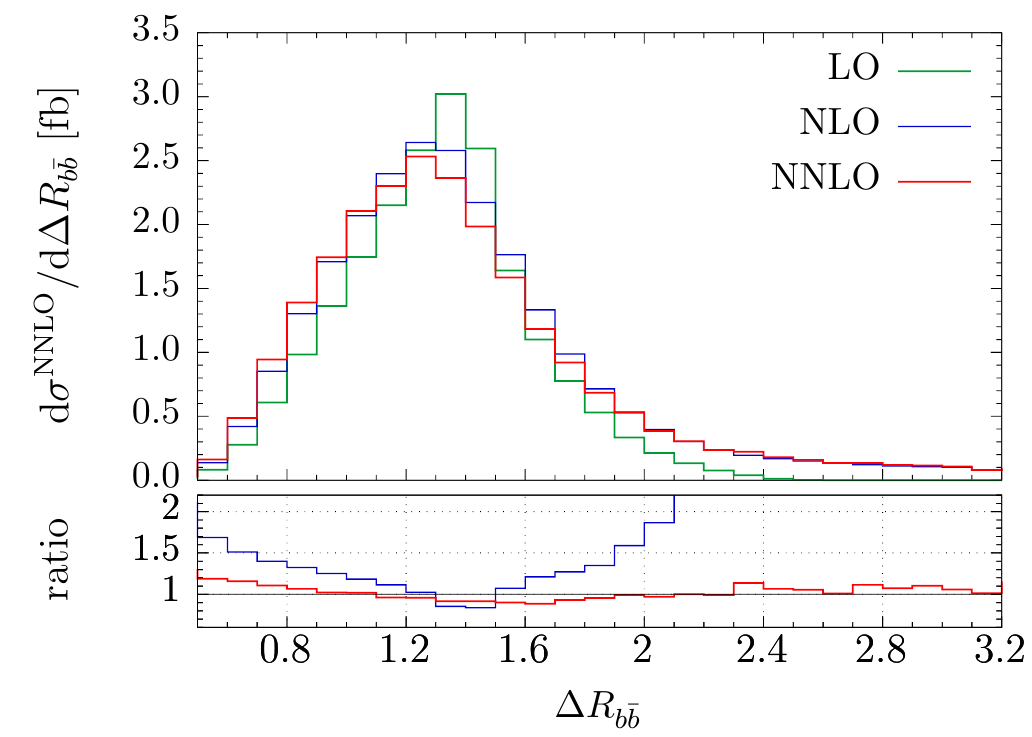}
\caption{The $\Delta R_{b \bar b}$
distribution of the two $b$-jets 
used to reconstruct  the Higgs boson. 
Left pane -- without the $p_\perp^{W}$ cut, right pane -- with 
the $p_{\perp}^{W} > 150~{\rm GeV}$ cut.  Lower panes -- ratios of NLO to LO and 
full NNLO to NLO distributions. See text for further details. }
\label{Rbb}
\end{figure}

It is also  interesting to study the angular
separation $\Delta R_{b \bar b} = \sqrt{\Delta \eta_{b \bar b}^2 + \Delta \phi_{b \bar b}^2}$  
of the $b$- and $\bar b$-jets
that are used to reconstruct the Higgs boson; the 
corresponding distributions 
without (left pane) and with (right pane)  the $p_{\perp,W}$ cut are shown in Fig.~\ref{Rbb}. 
The impact of the $W$ boson transverse momentum cut 
on the angular separation of the jets is dramatic,  
as the comparison of left and right panes shows.
The shift to lower values of $\Delta R_{b \bar b}$ is again expected, as imposing the $p_{\perp,W}$ cut selects boosted Higgs kinematics whose decay products are closer together. 
Both with and without the $p_{\perp,W}$ cut, the NLO corrections 
modify the shape of  $\Delta R_{b \bar b}$ distributions 
significantly, while the NNLO  corrections have a much smaller impact. 

\begin{figure}[t]
\centering
\includegraphics[width=0.485\textwidth]{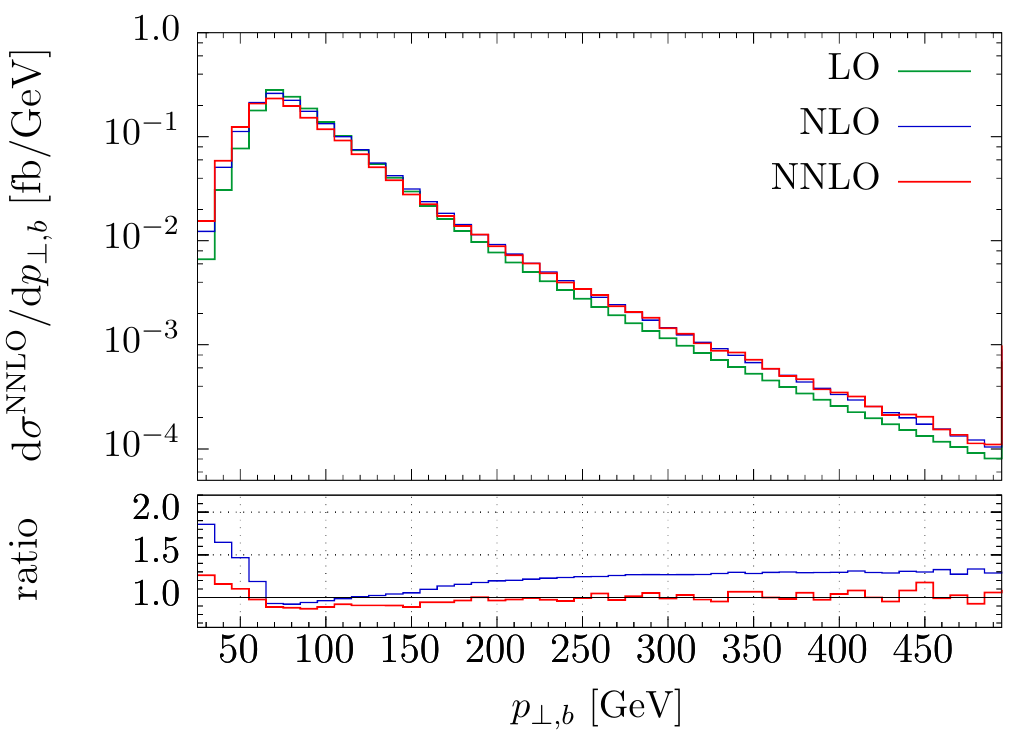}
~~
\includegraphics[width=0.485\textwidth]{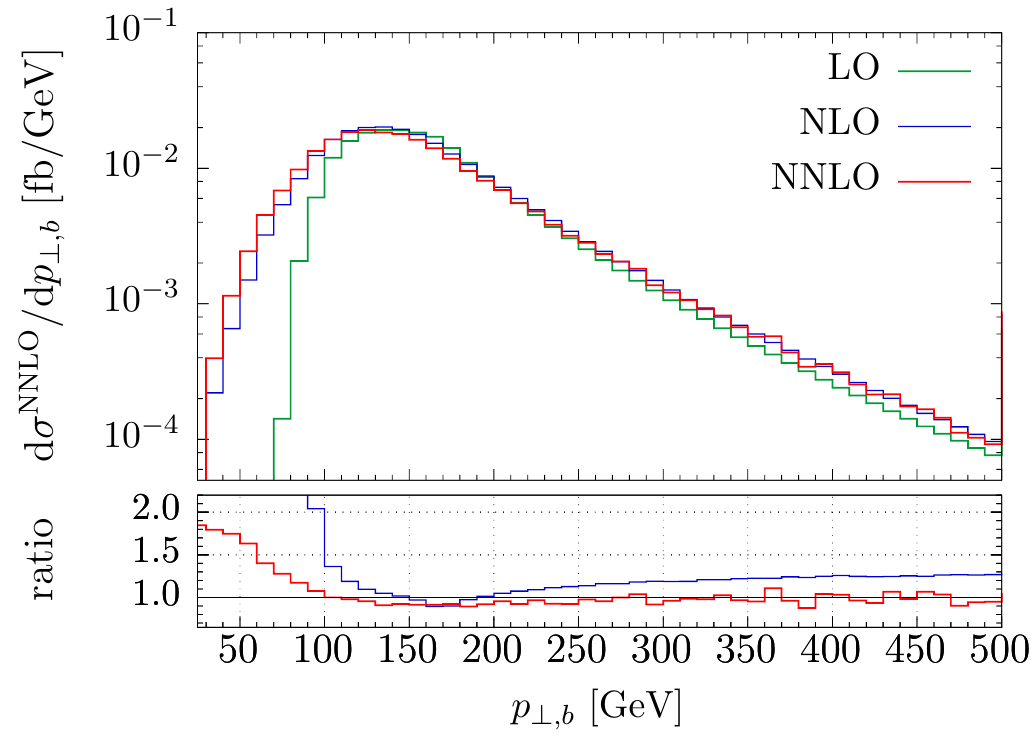}
\caption{The transverse momentum of the hardest $b$- or $\bar b$-jet.
Left pane -- without the $p_\perp^{W}$ cut, right pane -- with 
the $p_{\perp}^{W} > 150~{\rm GeV}$ cut.  
Lower panes -- ratios of NLO to LO and 
full NNLO to NLO distributions.
See text for further details.}
\label{bjetpt}
\end{figure}

Another  distribution 
that  is subject to large modifications if the cut on the vector boson 
transverse momentum is applied is the 
transverse momentum distribution of the hardest $b$-jet; it is shown in 
Fig.~\ref{bjetpt}.  In this case, 
 large radiative corrections appear 
below the value of the transverse momentum where the distribution reaches its maximum. 
If the $p_{\perp W}$ cut is not applied, large corrections at NLO 
are followed by moderate corrections at NNLO.  On the contrary, if  the 
$p_{\perp W}$ cut is in place, both the NLO and NNLO corrections are very large 
and perturbation theory does not appear to converge (see 
the right pane in  Fig.~\ref{bjetpt}).  Clearly, the situation 
is completely different at high values of $p_\perp^b$ where NNLO effects are relatively 
small and the NNLO/NLO $K$-factor is flat and close to one.

\begin{figure}[t]
\centering
\includegraphics[width=0.485\textwidth]{./plots/LONLONNLO_ptl.pdf}
~~
\includegraphics[width=0.485\textwidth]{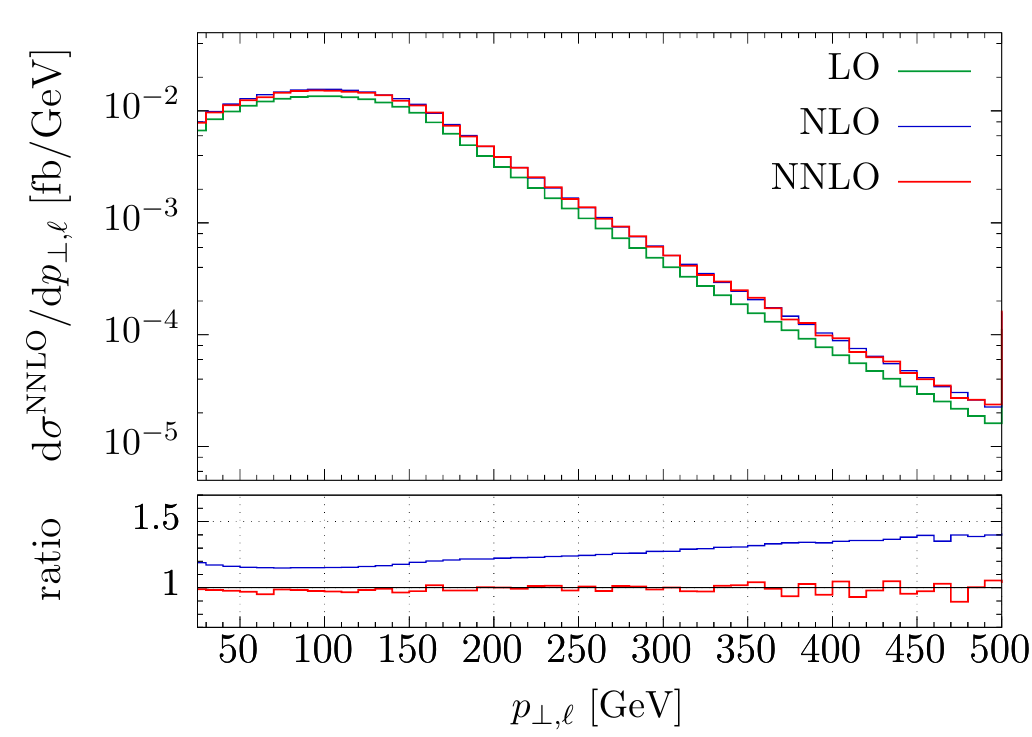}
\caption{The transverse momentum distribution of the charged lepton. 
Left pane -- without the $p_\perp^{W}$ cut, right pane -- with 
the $p_{\perp}^{W} > 150~{\rm GeV}$ cut.  
Lower panes -- ratios of NLO to LO and NNLO to NLO distributions. 
See text for further details.}
\label{ptl}
\end{figure}

As the last example, we show in Fig.~\ref{ptl} the transverse momentum distribution of the  
charged lepton that originates from the $W$ decay.  In this case, the 
cut on the $W$ boson transverse momentum has a significant impact on the shape 
of the  distribution, but the NLO and NNLO corrections to the two cases are 
very similar.  In particular, the NNLO corrections in both cases are relatively 
small and do not change the shape of the respective  NLO distributions.

\section{Comparison of fixed order and parton shower predictions}
\label{sec:partsh}
The goal of this Section is to compare fixed order QCD predictions for $pp \to
W(l \nu)H(b \bar b)$, described in the previous Section, with the
results obtained when parton showers are used to account for QCD radiation in
$H \to b \bar b$ decays, as typically done in many experimental analyses.  
We use the publicly available {\tt HWJ}
generator~\cite{NLOJPS} implemented in the {\tt POWHEG BOX}
framework~\cite{Nason:2004rx,Frixione:2007vw,Alioli:2010xd} to compute 
the process $p p \to W(l\nu)H+j$ at NLO QCD accuracy. 
In order to be as close as possible to the NNLO calculation, and
since the {\tt HWJ} generator allows it, we run it with the improved {\tt
  MiNLO} method~\cite{Hamilton:2012np,Hamilton:2012rf}.
This allows  observables that are inclusive in the
production of the color-neutral system, i.e. quantities in which the jet
is unresolved, to be computed with NLO QCD accuracy.
Thus, the difference between the NNLO fixed order calculation 
and the NLO parton shower simulation for the process $pp \to W(l \nu)H$ 
 is formally due only to
the missing two loop amplitudes in the {\tt HWJ} generator. 
 The decay of the Higgs boson to a $b \bar
b$ pair and an arbitrary number of gluons is instead simulated with a parton
shower using {\tt PYTHIA-8}~\cite{Sjostrand:2007gs} with the default
tune. Since we want to compare the parton shower results with a fixed order
calculation, we do not include any non-perturbative effects in the simulation,
i.e. the hadronization and the multi-parton interactions are switched off.
In the parton shower
simulation we reconstruct jets using the anti-$k_t$ algorithm~\cite{Cacciari:2008gp}, and
select $b$-jets according to Monte Carlo truth, 
in order to be as close as possible to experimental analyses. 
 Following Ref.~\cite{Ferrera:2017zex}
and the analysis in the previous Section, we use $R=0.5$ for the jet
radius. 

\begin{figure}[t]
\centering
\includegraphics[width=0.485\textwidth]{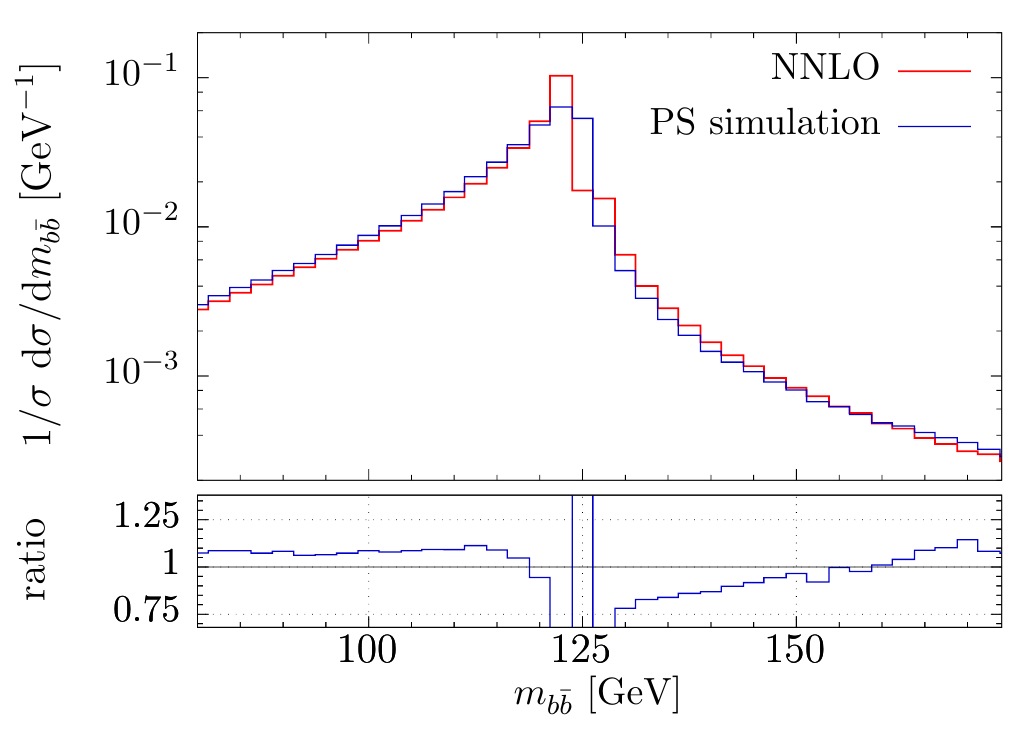}
~~
\includegraphics[width=0.485\textwidth]{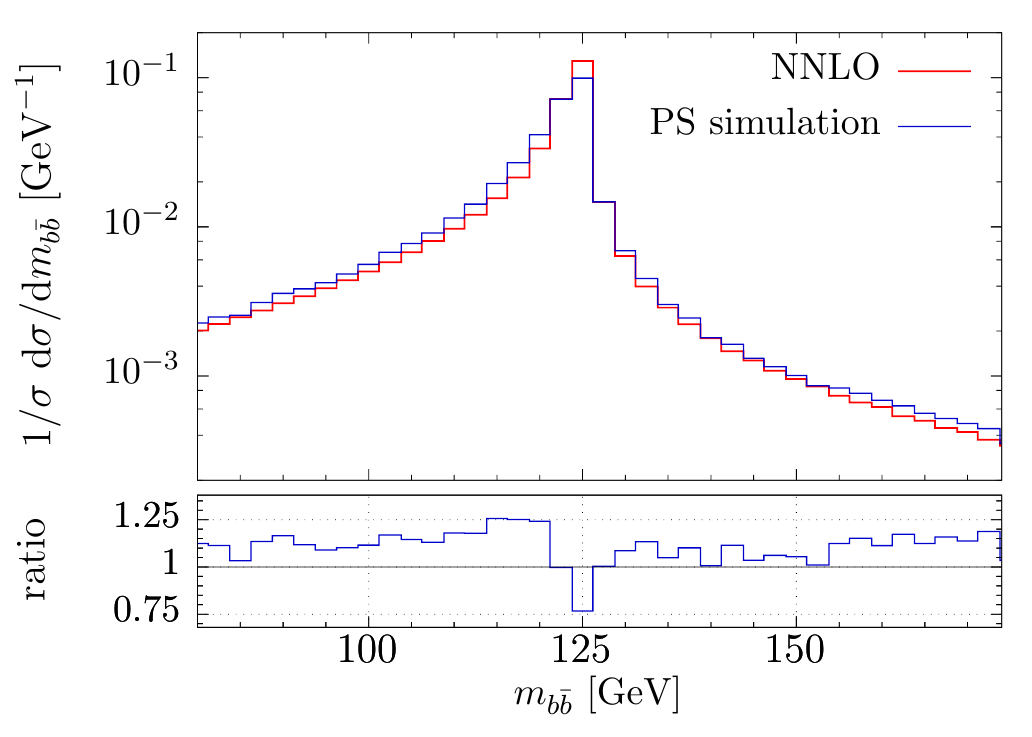}
\caption{Comparison of fixed order and parton shower predictions for the normalized 
invariant mass distribution of the two $b$-jets used to reconstruct the Higgs boson. Left pane --
  without the $p_\perp^{W}$ cut, right pane -- with the $p_\perp^{W} >
  150~{\rm GeV}$ cut.  Lower panes -- ratio of parton shower to fixed order predictions.
See text for further details.  }
\label{PaSh}
\end{figure}

As we have seen in the previous Section, radiative corrections to kinematic
distributions in the $pp \to WH(b \bar b)$ process exhibit non-trivial
patterns, partially because of selection criteria that are applied to final
state particles.  In particular, large effects are observed for values of the
$m_{b \bar b}$ invariant mass that are far from the value of the Higgs boson
mass, or for values of the transverse momenta of the $b \bar b$ system or the
leading $b$-jet that are below the cut on the transverse momentum of the
$W$ boson. All these kinematic regions have one thing in common -- they are not populated
{\it at all} if leading-order predictions are used. Hence, they require
additional QCD radiation either in the production process or in the decay of
the Higgs boson.

Moreover, some of these regions, e.g. $p_{\perp, b \bar b} \sim p_{\perp,W}^{\rm cut}$ or
hardest $p_{\perp, b} \to 0$, are close to kinematic boundaries
where parton showers
are known to accurately describe radiation effects.  Other regions and
observables, for example the case $m_{b \bar b} < m_H$ require a relatively
hard gluon emission and it is unclear \textit{a priori} if parton showers do a
good job in describing them.

\begin{figure}[t]
\centering
\includegraphics[width=0.485\textwidth]{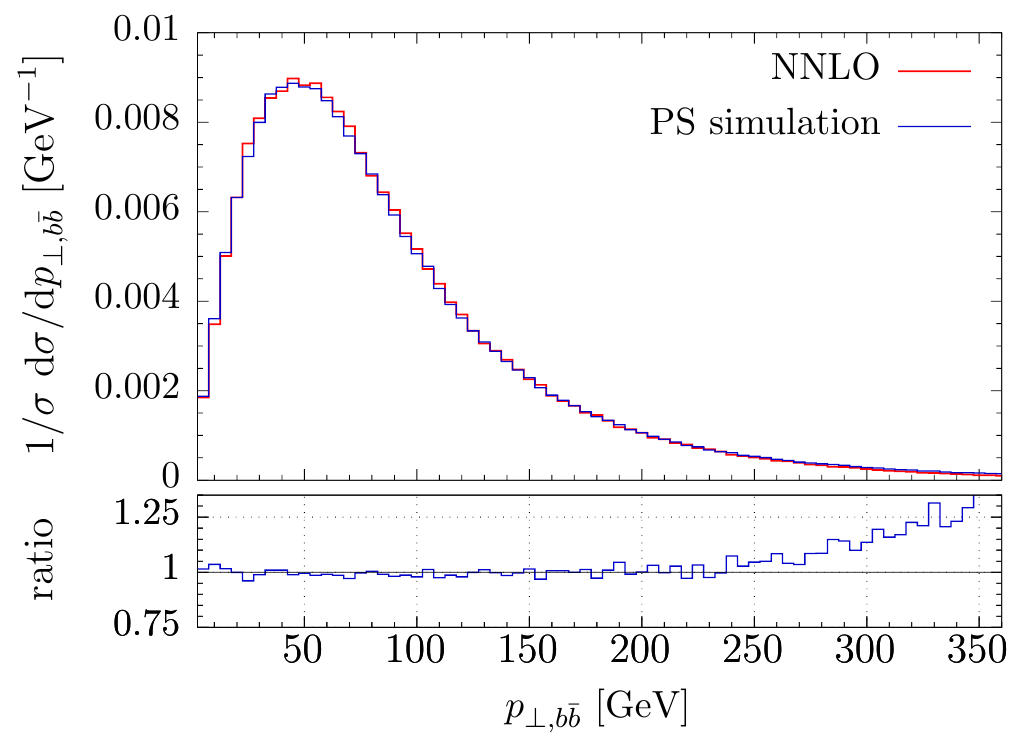}
~~
\includegraphics[width=0.485\textwidth]{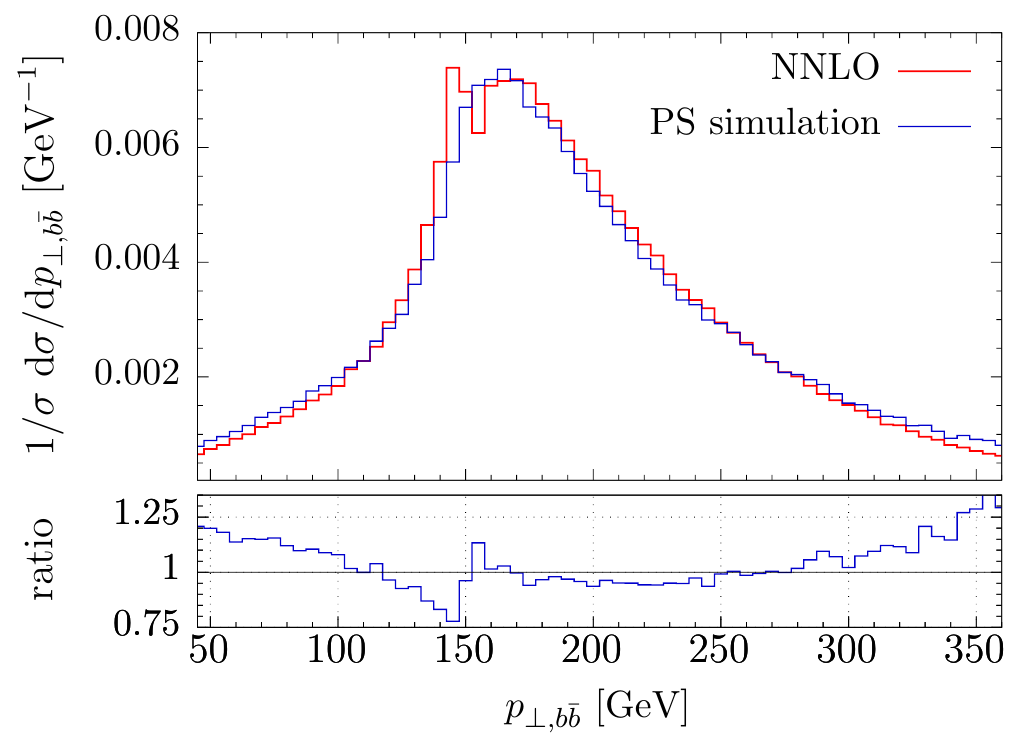}
\caption{Same as Fig.~\ref{PaSh} but for the transverse momentum of the $b\bar b$ system that
is used to reconstruct the Higgs boson. See text for further details.}
\label{PaSh1}
\end{figure}

\begin{figure}[t]
\centering
\includegraphics[width=0.485\textwidth]{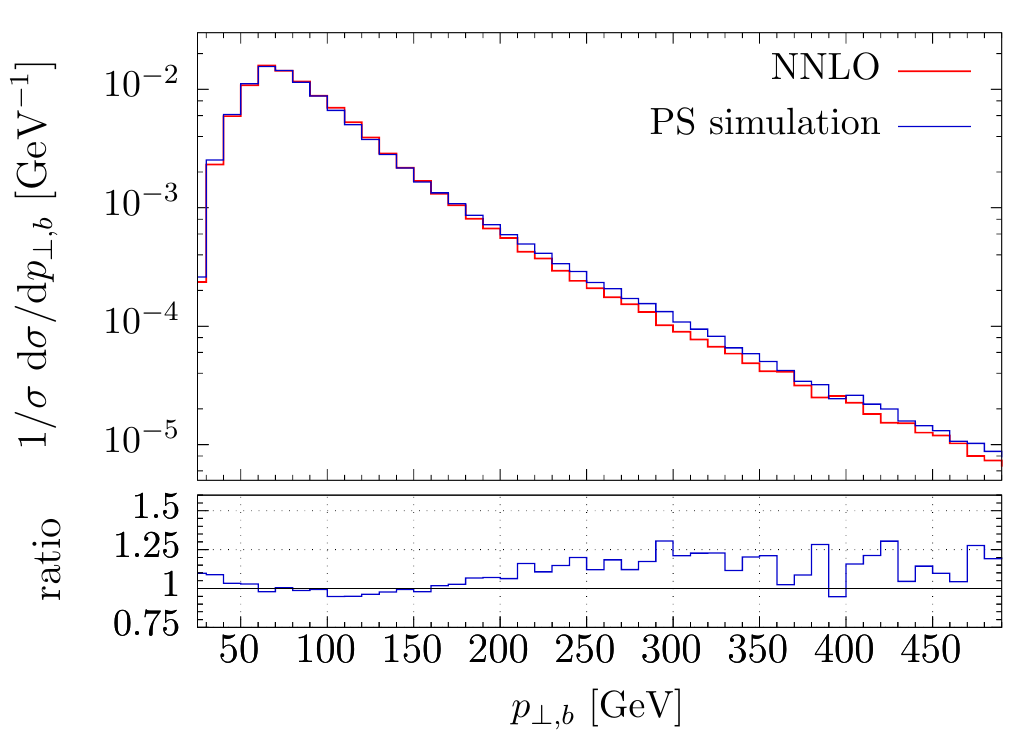}
~~
\includegraphics[width=0.485\textwidth]{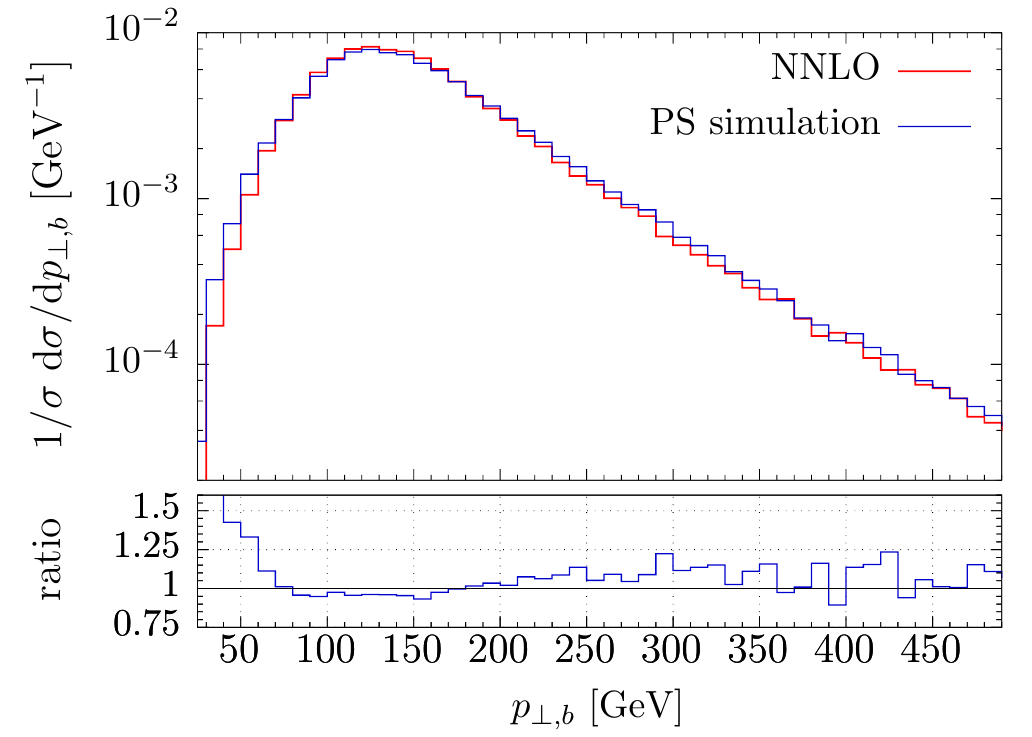}
\caption{Same as Fig.~\ref{PaSh} but for the hardest $b$ (or $\bar b$) jet.
See text for further details.}
\label{PaSh2}
\end{figure}

As in the previous Section, we study the $b$ and $\bar b$ jets whose
invariant mass $m_{b\bar b}$ is closest to the Higgs mass. We show a comparison of the NNLO
and parton shower predictions for the $m_{b\bar b}$ distribution in Fig.~\ref{PaSh}, 
for the transverse momentum distribution
of the $b\bar b$ system in Fig.~\ref{PaSh1}, and for the hardest $b$ (or $\bar b$) jet
$p_\perp$ distribution in Fig.~\ref{PaSh2}. In all of these cases, the distributions
are normalized to their inclusive result so that their shapes can be compared.  
However, we note that,
while the fixed order and parton shower results use the same jet radius, the
former makes use of the flavor-$k_t$ jet algorithm while the latter uses the
standard anti-$k_t$ algorithm, and therefore the comparison between the two is not
straightforward. We will return to this point at the end of this Section.

For the $m_{b \bar b}$ distribution, we observe that the parton shower does quite
a good job in describing the NNLO corrections, although it
predicts more events at both low and high  values of $m_{b \bar b}$.
Interestingly, the parton shower smears the peak at $m_{b\bar b}=m_H$ more
 significantly in the case where the
$p_{\perp,W}$ cut is not applied. When this cut is imposed, the parton shower predicts 
fewer events at the peak but the smearing effect is not as dramatic.

Turning to the $p_{\perp, b \bar b}$ distribution, we observe that the parton shower is
able to describe the NNLO distributions quite well. When the $p_{\perp}^{W}$ cut is not imposed,
the parton shower prediction is in excellent agreement with the fixed order one, except in
the very high transverse momentum region. 
However,  there is a 
difference at low $p_{\perp, b \bar b}$ if the $p_\perp^W$ cut is applied,
with the parton shower predicting more events in this region than the
fixed order calculation.  As expected, the parton shower also removes the
Sudakov shoulder in this distribution that was observed in both the
approximate and the full NNLO distributions.

Next, in Fig.~\ref{PaSh2} we show the $p_\perp$ distribution of the $b$- (or $\bar b$-) jet 
with largest transverse momentum. Without the cut on $p_\perp^W$, the NNLO and shower results
are similar, although the latter predicts slightly more events at large $p_\perp$. 
On the other hand, if the cut $p_\perp^W>150~{\rm GeV}$
is imposed, the fixed order and shower calculations deviate significantly at small $p_\perp$. 
Large shower effects in this region are expected, since as we have shown in
 Section~\ref{sect:fullpheno}, the fixed order predictions are not reliable here.

\begin{figure}[t]
\centering
\includegraphics[width=0.485\textwidth]{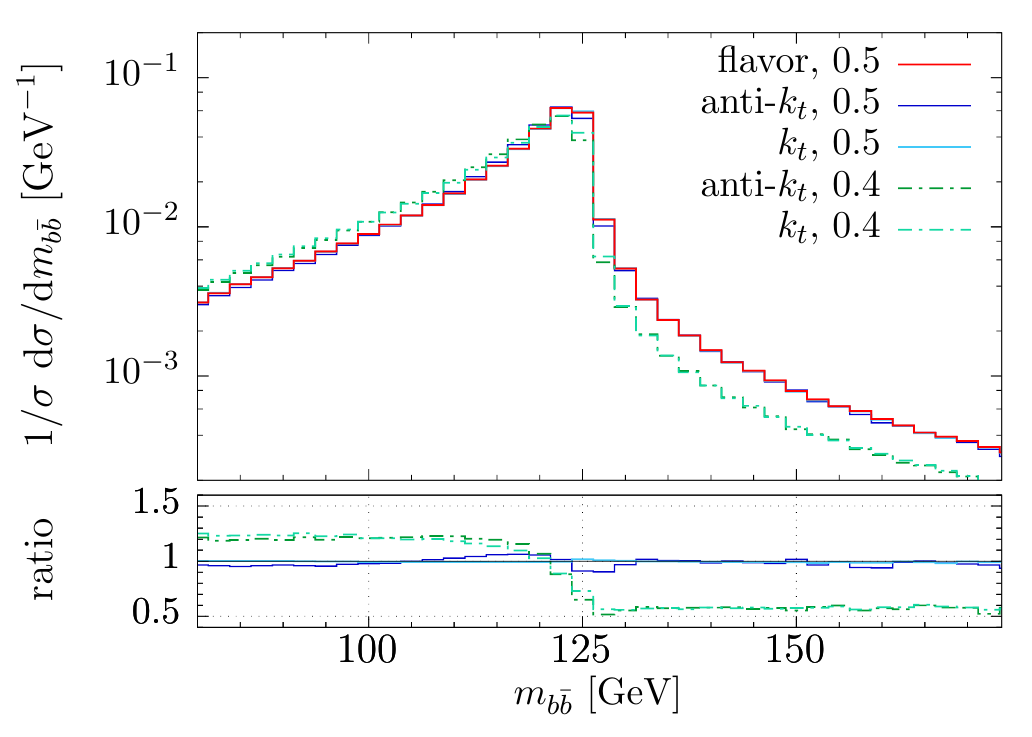}
~~
\includegraphics[width=0.485\textwidth]{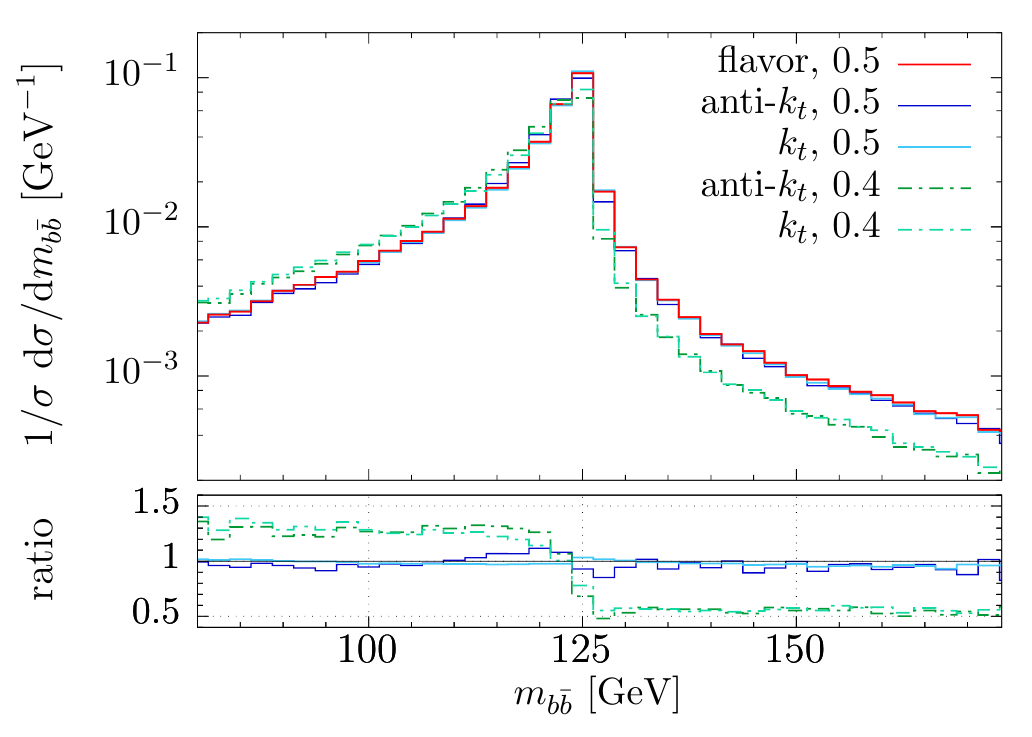}
\caption{The invariant mass of a $b$-jet and a $\bar b$-jet that best approximates the Higgs 
boson mass, obtained from parton shower simulations with different jet algorithms and radii.
Left pane -- without the $p_\perp^W$ cut, right pane -- with the $p_\perp^W > 150~\rm{GeV}$ cut. 
Lower panes -- ratio of results for the $k_t$ and anti-$k_t$
 jet algorithms with $R=\{0.4,0.5\}$ to the result for the flavor-$k_t$ jet algorithm~
with $R=0.5$. See text for further details. }
\label{jetcomp_mbb}
\end{figure}

\begin{figure}[t]
\centering
\includegraphics[width=0.485\textwidth]{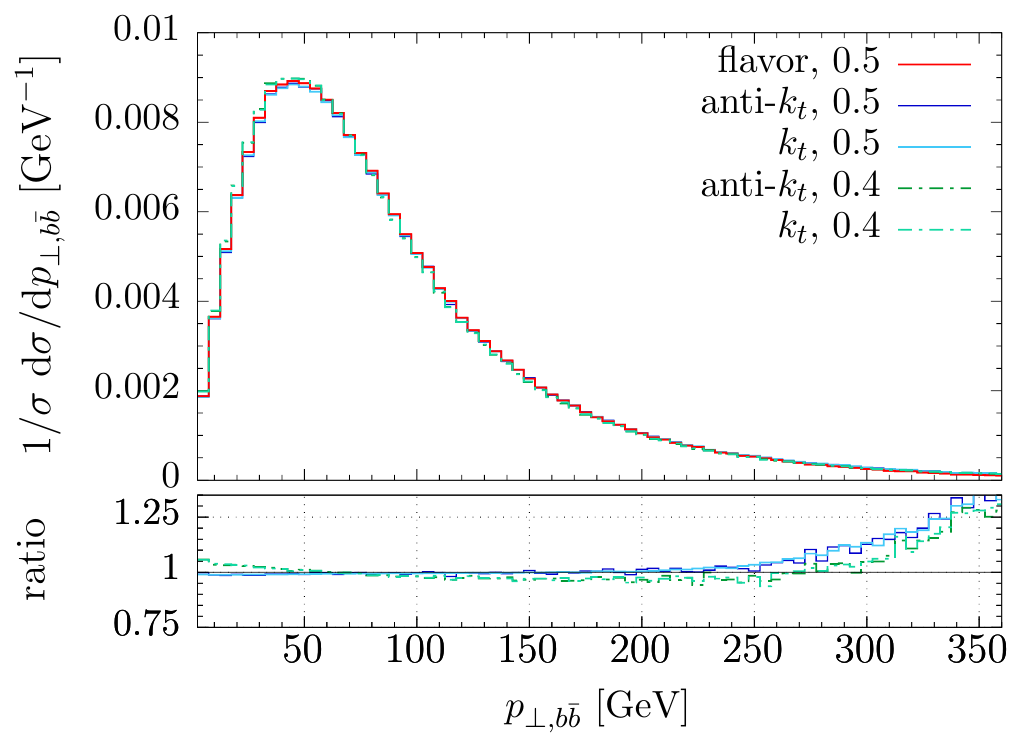}
~~
\includegraphics[width=0.485\textwidth]{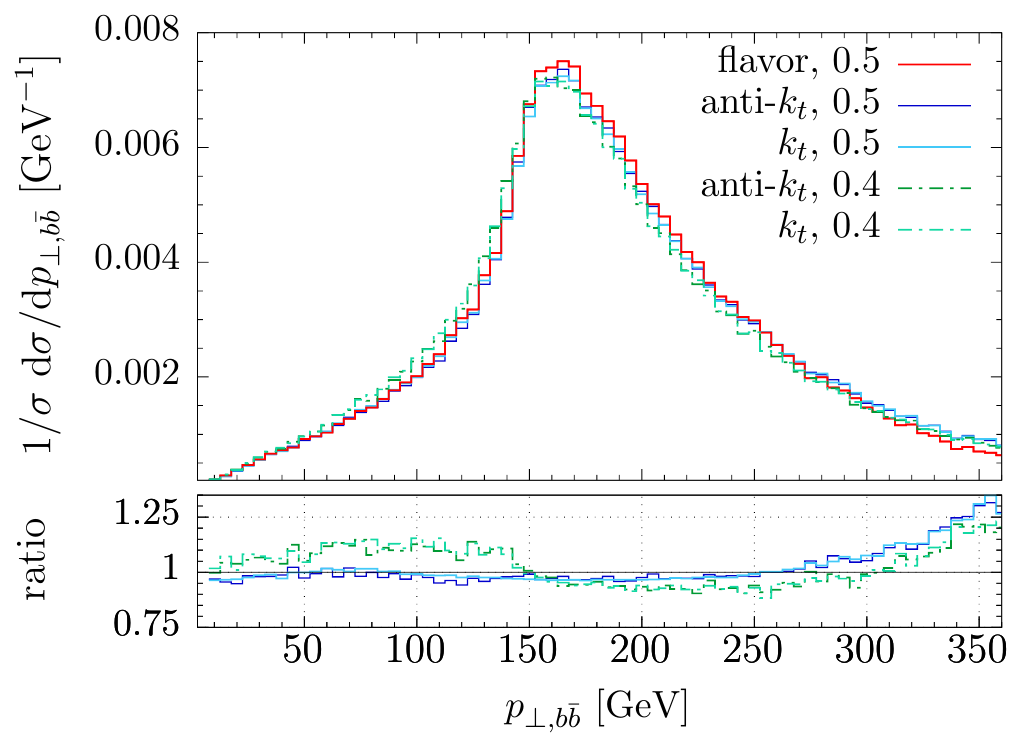}
\caption{Same as Fig.~\ref{jetcomp_mbb} but for the transverse momentum of the $b\bar b$
system that is used to reconstruct the Higgs boson.  See text for further details. }
\label{jetcomp_ptbb}
\end{figure}

Given the different jet algorithms used in the fixed order and parton shower
calculations, it is interesting to investigate to what extent the details of the jet definition
affect these results. In Figs.~\ref{jetcomp_mbb} and~\ref{jetcomp_ptbb}, 
we show the invariant mass $m_{b\bar b}$ and transverse momentum distribution $p_{\perp,b\bar b}$,
obtained from the parton shower simulation for different choices
of the jet algorithm and radius. We compare
the flavor-$k_t$ jet algorithm~\cite{Banfi:2006hf} 
with both the $k_t$~\cite{kt} and anti-$k_t$~\cite{Cacciari:2008gp} algorithms. 
For the invariant mass distribution, Fig.~\ref{jetcomp_mbb} shows that both with and
without the $p_\perp^W>150~{\rm GeV}$ cut the result is quite insensitive to the recombination
algorithm, and it only depends on the choice of the jet radius: smaller values of $R$
lead to more events below the Higgs peak. For the $p_{\perp,b\bar b}$ on the other hand, 
Fig.~\ref{jetcomp_ptbb} shows that without the $p_\perp^W$ cut all jet algorithms and radii lead
to the same result, apart from the high $p_{\perp,b\bar b}$ tail where the flavor-$k_t$ 
jet algorithm~\cite{Banfi:2006hf} predicts fewer events compared to the $k_t$ and anti-$k_t$ cases.
With the additional $p_\perp^W > 150~{\rm GeV}$ cut, a qualitative dependence on the
jet radius similar to the one seen in the $m_{b\bar b}$ distribution is observed: 
smaller values of $R$ lead to a softer spectrum. 

\section{Conclusions} 
\label{sect:concl}

In this paper we
presented  a computation of the NNLO QCD corrections to 
the associated production  of the Higgs boson $pp \to WH$ at the LHC. We considered 
the $H\to b\bar b$ decay of the 
Higgs boson and included radiative corrections to this decay 
through NNLO in perturbative QCD. 

We pointed out an interesting contribution to Higgs decay to $b \bar b$ pairs 
that was ignored in previous {\it fully-differential} NNLO QCD computations 
to this process.  This contribution is infrared-sensitive even after standard jets 
algorithms are applied and understanding it necessitates the computation of fully-differential NNLO
corrections to the $H \to b \bar b$ decay with massive bottom quarks.  Although we argued
that its numerical importance should be small, a more refined analysis is required 
to properly quantify these effects.

We found a number of kinematic distributions in the $pp\to W(l\nu)H(b\bar b)$ process  that receive 
large perturbative corrections if certain cuts on the final state, and especially 
a cut on the transverse momentum of the $W$ boson, are applied.  These findings 
are in accord with an earlier discussion given in Ref.~\cite{Ferrera:2017zex}.

We compared fixed order predictions  for the $pp \to W(l\nu)H(b \bar b)$ process with calculations 
where a parton shower is used to describe QCD radiation  in $H \to b \bar b$ decay. Parton showers 
confirm the existence of large effects observed in fixed order computations.
   Since at the moment fixed order NNLO QCD computations 
for $H \to b \bar b$ are performed for massless $b$-quarks, one has to use specially-tailored 
jet algorithms to describe flavored jets in fixed order computations~\cite{Banfi:2006hf}. 
Although we showed in Section~\ref{sec:partsh} that the results are largely insensitive to
the jet recombination algorithm, it would be interesting to repeat the fixed order
studies reported here for the setup used in experimental analyses. 
This requires the computation of the fully-differential decay $H \to b \bar b$ 
through NNLO QCD  keeping the full dependence on the $b$-quark mass. It would also be interesting
to compare fixed order predictions to more advanced parton shower implementations, as
described e.g. 
in Refs.~\cite{Richardson:nlodecay,Hoeche:2014aia,Alioli:2015toa,Astill:2016hpa}.
We leave these investigations for future work.

 {\bf Acknowledgments} 
We would like to thank  Munich Institute for Astronomy and Particle Physics (MIAPP) 
for  hospitality and partial support during the programs 
\emph{Automated, Resummed and Effective} and \emph{Mathematics and physics of scattering amplitudes}.
The research of K.M. and R.R. is  partially supported by BMBF grant 05H15VKCCA. 
The research of F.C. was supported in part by the ERC starting grant 637019 ``MathAm''. 
We are grateful to G.~Salam for interesting discussions and for providing us with
a private implementation of the flavor-$k_t$ jet algorithm~\cite{Banfi:2006hf}.  We are indebted  
to F.~Tramontano for discussions and for his help in comparison with the results 
of Ref.~\cite{Ferrera:2017zex}.

\end{document}